\documentclass[%
reprint,
superscriptaddress,
amsmath,amssymb,
aps,
prx,
]{revtex4-2}

\usepackage{graphicx}
\usepackage{dcolumn}
\usepackage{bm}
\usepackage{hyperref}

\usepackage{txfonts}
\usepackage{braket}
\usepackage{comment}
\usepackage{color}

\begin{document}

\newcommand{\CModel}{ALG}
\newcommand\red[1]{\textcolor{red}{#1}}
\newcommand\blue[1]{\textcolor{blue}{#1}}
\newcommand{\eff}{\mathrm{eff}}

\title{Activity-induced phase transition in a quantum many-body system}

\author{Kyosuke Adachi}
\affiliation{Nonequilibrium Physics of Living Matter RIKEN Hakubi Research Team, RIKEN Center for Biosystems Dynamics Research, 2-2-3 Minatojima-minamimachi, Chuo-ku, Kobe 650-0047, Japan}
\affiliation{RIKEN Interdisciplinary Theoretical and Mathematical Sciences Program, 2-1 Hirosawa, Wako 351-0198, Japan}

\author{Kazuaki Takasan}
\affiliation{Department of Physics, University of California, Berkeley, California 94720, USA}
\affiliation{
Materials Sciences Division, Lawrence Berkeley National Laboratory, Berkeley, California 94720, USA
}

\author{Kyogo Kawaguchi}
\affiliation{Nonequilibrium Physics of Living Matter RIKEN Hakubi Research Team, RIKEN Center for Biosystems Dynamics Research, 2-2-3 Minatojima-minamimachi, Chuo-ku, Kobe 650-0047, Japan}
\affiliation{RIKEN Cluster for Pioneering Research, 2-2-3 Minatojima-minamimachi, Chuo-ku, Kobe 650-0047, Japan}
\affiliation{Universal Biology Institute, The University of Tokyo, Bunkyo-ku, Tokyo 113-0033, Japan}

\date{\today}

\begin{abstract}
A crowd of nonequilibrium entities can show phase transition behaviors that are prohibited in conventional equilibrium setups.
An interesting question is whether similar activity-driven phase transitions also occur in pure quantum systems.
Here we introduce a minimally simple quantum many-body model that undergoes quantum phase transitions induced by non-Hermiticity.
The model is based on a classical anisotropic lattice gas model that undergoes motility-induced phase separation (MIPS), and the quantum phase diagram includes other active phases such as the flocking phase.
The quantum phase transitions, which in principle can be tested in ultracold atom experiments, is also identified as the transitions of dynamical paths in the classical kinetics upon the application of biasing fields. This approach sheds light on the useful connection between classical nonequilibrium kinetics and non-Hermitian quantum physics.
\end{abstract}

\maketitle

\section{Introduction}

The collective dynamics of active or self-driven components can lead to phase transitions and pattern formations that are prohibited in equilibrium systems~\cite{Marchetti2013}.
Recent works have shown the properties of materials such as surface flow~\cite{nash2015}, odd responses~\cite{banerjee2017}, and anomalous topological defect dynamics~\cite{sanchez2012} that can be realized by introducing activity into the design.
In addition to the application in biophysical examples~\cite{prost2015}, combining the understandings of active systems with a broader range of models in condensed matter should bring progress not only in nonequilibrium physics but also in material science~\cite{needleman2017}.

Although the scope of active matter has greatly widened in the past years~\cite{Gompper2020}, its quantum analog has so far not been explicitly proposed.
Part of the reason is that the corresponding quantum system must be open (i.e., non-Hermitian), which is challenging to examine both in experiment and theory compared with closed (i.e., Hermitian) systems.
In recent years, however, advances in atomic-molecular-optical experiments have allowed precise control over open quantum systems~\cite{Muller2012, Daley2014, Schafer2020, ashida2020}, encouraging the exploration of nonequilibrium physics in various courses including topological phases~\cite{Hu2011, Esaki2011, Shen2018, Gong2018, Yao2018, Kawabata2019, ozawa2019topological, Li2020} and quantum critical phenomena~\cite{Ashida2017, Nakagawa2018, Hamazaki2019, Kawabata2021} in non-Hermitian setups.
We are therefore in position to ask whether there exist new phases of matter induced by activity (i.e., dissipative terms that can be interpreted as self-driving) in quantum many-body systems, and if so, how they can be realized in experiments.

The formal connection between classical stochastic dynamics and quantum mechanics has been extensively utilized in the field of statistical physics.
For example, tools such as the Bethe ansatz have provided useful in solving models of classical nonequilibrium dynamics~\cite{gwa1992bethe}, even though the corresponding quantum Hamiltonian becomes non-Hermitian.
Simulation algorithms such as population dynamics, where selection processes are added on top of classical stochastic simulations, have been used to study the ground state of quantum many-body systems (i.e., quantum Monte Carlo)~\cite{mcmillan1965ground}.

One of the simplest models of phase transition in active matter is the exclusion process with uni-axial activity~\cite{Kourbane2018}.
In this model, the particles undergo exclusive random walk with uni-axially biased hopping depending on their internal degree of freedom.
This model undergoes anisotropic phase separation upon increasing the strength of biased-hopping (i.e., self-driven motility), which could be thought of as an example of motility-induced phase separation (MIPS)~\cite{Cates2015}.
MIPS has been observed in Brownian particle~\cite{Fily2012} and lattice~\cite{Thompson2011} simulations as well as in experiments involving artificial~\cite{Buttinoni2013} and biological~\cite{Liu2019} materials.
Although the basic mechanism of MIPS is seemingly simple (i.e., accumulation of particles at high-density regions due to the slowing down of self-propelled motion), the components that cause the phase separation behavior~\cite{Solon2018,Solon2018b} and the critical properties of the phase transitions~\cite{Siebert2018,Partridge2019} are still under active discussion, and may depend on the details of the model~\cite{Tjhung2018,Shi2020}.

Similar models with uni-axial biased motion have been considered as the driven lattice gas, where phase behaviors and spatio-temporal correlations have been shown to drastically change due to the anisotropy~\cite{Schmittmann1995,Marro1999}.
Anisotropic models of active lattice gas have also appeared in the context of flocking~\cite{Solon2013,Solon2015c}, where a macroscopic number of particles collectively move in one direction~\cite{Vicsek2012}.
Yet, despite the simplicity of these models, the relation between the activity-induced phase transitions and the anomalous behaviors owing to the spatial anisotropy has not been extensively discussed.

In this work, we introduce a quantum many-body model on a lattice with an analog of uni-axial activity.
We show that this minimally simple model embeds a classical interacting particle model within its parameter space, where the self-driving property of the particles is encoded in the non-Hermiticity of the Hamiltonian.
We find that the embedded classical model undergoes anisotropic MIPS with interesting properties even in the seemingly trivial homogeneous phase, and the phase separation behavior appears as the property of the ground-state in the quantum model.
By investigating the phase diagram via Monte Carlo (MC) simulation, we find that the quantum model exhibits flocking and microphase-separated phases that do not appear in the embedded classical model, and further show the relation of these phases to the dynamical phases that have been discussed in the context of glassy systems~\cite{garrahan2007} and classical active matter~\cite{Whitelam2018,Nemoto2019,Tociu2019}.
Finally, we discuss the possibility of implementing the model in an ultracold atomic gas experiment.

\section{Non-Hermitian hard-core bosons and classical active lattice gas}

\begin{figure}[t]
\centering
\includegraphics[scale=0.8]{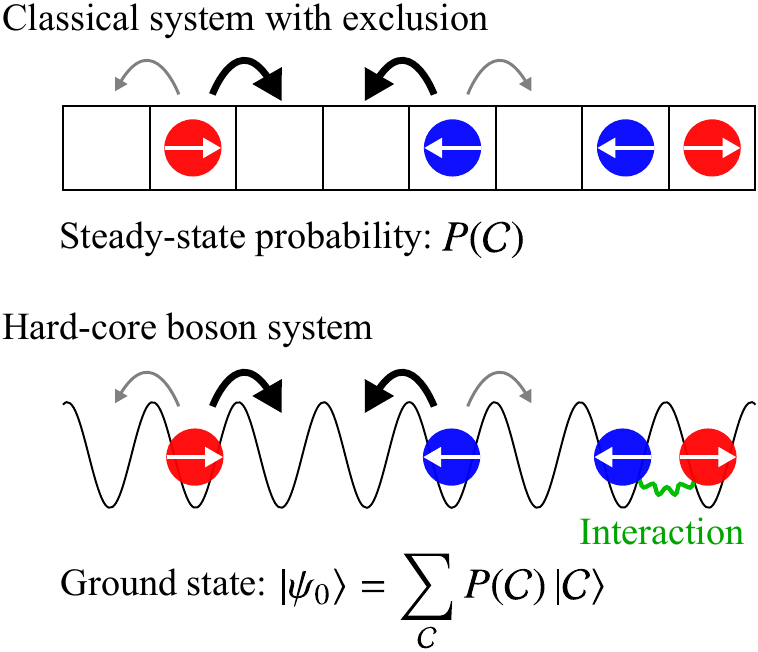}
\caption{Correspondence between a classical model of active matter and a quantum model of hard-core bosons.
In the classical model, particles stochastically move with asymmetric hopping rates.
In the quantum model, hard-core bosons asymmetrically hop due to the non-Hermitian terms in the Hamiltonian, and also feel nearest-neighbor interactions additional to the hard-core repulsion.
$\ket{\mathcal{C}}$ is the Fock-space basis corresponding to the microscopic configuration $\mathcal{C}$.}
\label{Fig:Scheme}
\end{figure}

The model we study here (Fig.~\ref{Fig:Scheme}) involves quantum hard-core bosons with ``spin'' $s$ ($= \pm 1$) in a $L_x \times L_y$ rectangular lattice with periodic boundary condition (PBC):
\begin{align}
& H(J,\varepsilon,U_1,U_2,h) = - J \sum_{\braket{i, j}, s} ( a_{i, s}^\dag a_{j, s} + a_{j, s}^\dag a_{i, s} ) \nonumber \\
& - \varepsilon J \sum_{i, s} s ( a_{i, s}^\dag a_{i - \hat{x}, s} - a_{i, s}^\dag a_{i + \hat{x}, s} ) - h \sum_{i, s} a_{i, s}^\dag a_{i, -s} \nonumber \\
& - U_1 \sum_{\braket{i, j}} \hat{n}_i \hat{n}_j - U_2 \sum_i \hat{m}_i (\hat{n}_{i + \hat{x}} - \hat{n}_{i - \hat{x}}) + ( 4 J + h ) N,
\label{Eq:Hamiltonian}
\end{align}
where $\hat{n}_{i,s}$ $:= a_{i, s}^\dag a_{i, s}$ is the local density of particles with spin $s$, $\hat{n}_i$ $:= \hat{n}_{i, +} + \hat{n}_{i, -}$, and $\hat{m}_i$ $:= \hat{n}_{i, +} - \hat{n}_{i, -}$.
$\hat{x}$ is the unit horizontal translation, and $N$ is the fixed total number of particles. 
The second term in \eqref{Eq:Hamiltonian} describes the spin-dependent asymmetric hopping ($J > 0$ and $-1 \leq \varepsilon \leq 1$), which is non-Hermitian for $\varepsilon \neq 0$.
The fourth and fifth terms represent the spin-independent and dependent nearest-neighbor interactions, respectively, with their general form discussed in Appendix~\ref{Sec:genq}.
We take $h > 0$ and consider a partial Fock space where multiple particles cannot occupy a single site regardless of their spins.

The physical interpretation of a non-Hermitian quantum system is ambiguous due to the complex energy spectrum.
For the case of \eqref{Eq:Hamiltonian}, however, its eigenvalue with the smallest real part is unique and real (which we call $E_0$), and the corresponding eigenstate can be taken to have all its elements real and positive (which we denote as $\ket{\psi_0}$).
This is due to the Perron-Frobenius theorem, which can be applied since the off-diagonal elements of $H$ in the Fock-space representation are all real and non-positive.
In this work, we focus on how the ground state $\ket{\psi_0}$ (with ground state energy $E_0$) changes according to the change of parameters in $H$.
Throughout this paper, we set $\hbar = 1$, so that energy has the dimension of the inverse of time.

\section{Anisotropic active lattice gas}

\begin{figure*}[t]
\centering
\includegraphics[scale=0.8]{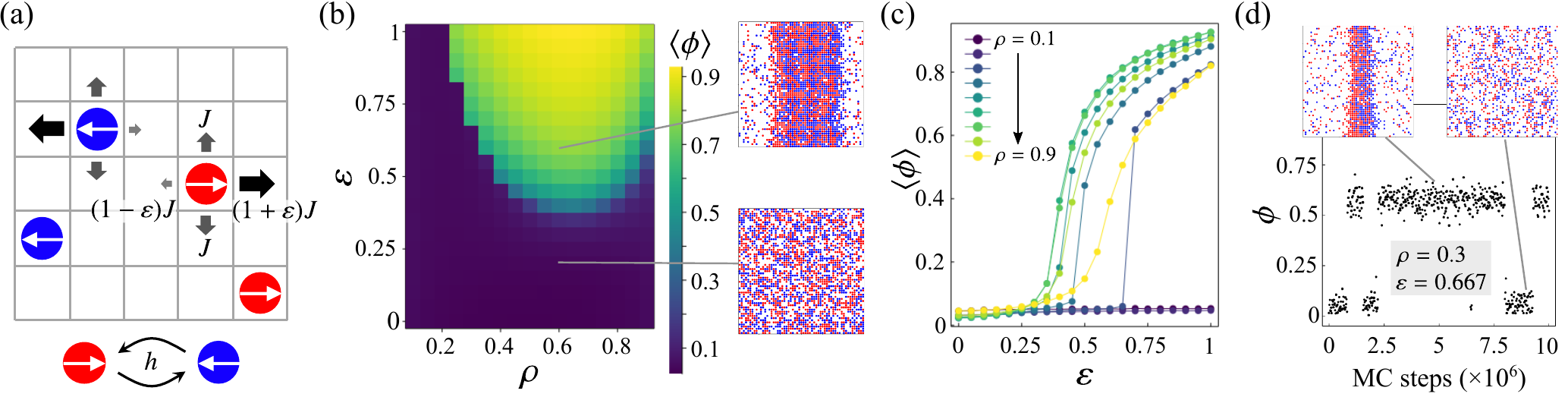}
\caption{(a) Anisotropic ALG.
Each particle with spin hops to the nearest-neighbor site with the spin-dependent rate or flips its spin with the rate $h$.
(b) Heatmap of the steady-state order parameter $\braket{\phi}$ in the $\rho$-$\varepsilon$ plane.
Typical configurations of the PS and homogeneous states are also shown.
(c) $\varepsilon$-dependence of $\braket{\phi}$ for different values of $\rho$, which is another plot of (b).
(d) Time-dependence of $\phi$ at a discontinuous transition point ($\rho = 0.3$ and $\varepsilon = 0.667$) with typical snapshots of bistable PS and homogeneous states.
For (b), (c), and (d), we set $L_x = L_y = 60$.}
\label{Fig:CMIPS}
\end{figure*}

Within the parameter space of \eqref{Eq:Hamiltonian}, there is a special subspace defined by $U_1 = 2J$ and $U_2 = \varepsilon J$, where the Hamiltonian can be mapped~\cite{Doi1976a,Peliti1985} to the transition rate matrix of an active lattice gas model (\CModel{}) (see Appendix~\ref{App:Mapping}).
The \CModel{} here is an $N$-particle model where the particles are exclusively hopping within the $L_x \times L_y$ rectangular lattice with the PBC [Fig.~\ref{Fig:CMIPS}(a)].
Each particle has a spin $s$ ($= \pm 1$) as its internal variable, which sets the rate of asymmetric hopping in the $x$-direction as $(1 + \varepsilon s) J$ and $(1 - \varepsilon s) J$ for the positive and negative directions, respectively.
The $y$-directional hopping rate is $J$, the spin flipping rate is $h$, and we define the density as $\rho := N / (L_x L_y)$ ($0 < \rho < 1$).

Before considering the full quantum model~\eqref{Eq:Hamiltonian}, we study static and dynamical properties of the ALG as an anisotropic active matter model.
In the following, we set $h = 0.025J$ in MC simulations (see Appendix~\ref{App:MC}).
As we increase $\varepsilon$, the ALG shows a phase transition from the homogeneous state to the phase-separated (PS) state [see the typical configurations in Fig.~\ref{Fig:CMIPS}(b)], where the particles moving in the $+x$ or $-x$ direction are blocked by others moving in the opposite direction.
Similar types of phase transitions have been discussed in two-species driven lattice gas models~\cite{Schmittmann1992,Bassler1993,Foster1994,Korniss1997} and recently regarded as a MIPS transition in an ALG~\cite{Kourbane2018}.
We define the order parameter for the PS state as $\braket{\phi} := \braket{|\sum_j \exp (- 2 \pi i x_j / L_x) n (\bm{r}_j)|} \sin (\pi / L_x) / [\sin ( \pi \rho ) L_y]$~\cite{Leung1991,Wang1996}, where $\bm{r}_j$ $[:= (x_j, y_j)]$ and $n (\bm{r}_j)$ are the spatial coordinate and occupancy of the site $j$, respectively, $\braket{\cdots}$ is the ensemble average in the steady-state, and $\braket{\phi} = 1$ for the fully PS state.
For $L_x = L_y = 60$, we obtain the $\rho$-$\varepsilon$ phase diagram as a heatmap of $\braket{\phi}$ [Fig.~\ref{Fig:CMIPS}(b)].
The $\varepsilon$-dependence of $\braket{\phi}$ [Fig.~\ref{Fig:CMIPS}(c)] and the bistability at the transition point [Fig.~\ref{Fig:CMIPS}(d)] suggest that the transition is discontinuous for low density ($\rho \lesssim 0.4$), as observed in similar models~\cite{Korniss1997}, though further investigation is needed to clarify whether the transition is still discontinuous in the thermodynamic limit.

\subsection{Long-range correlation in the homogeneous state}

\begin{figure}[t]
\centering
\includegraphics[scale=0.8]{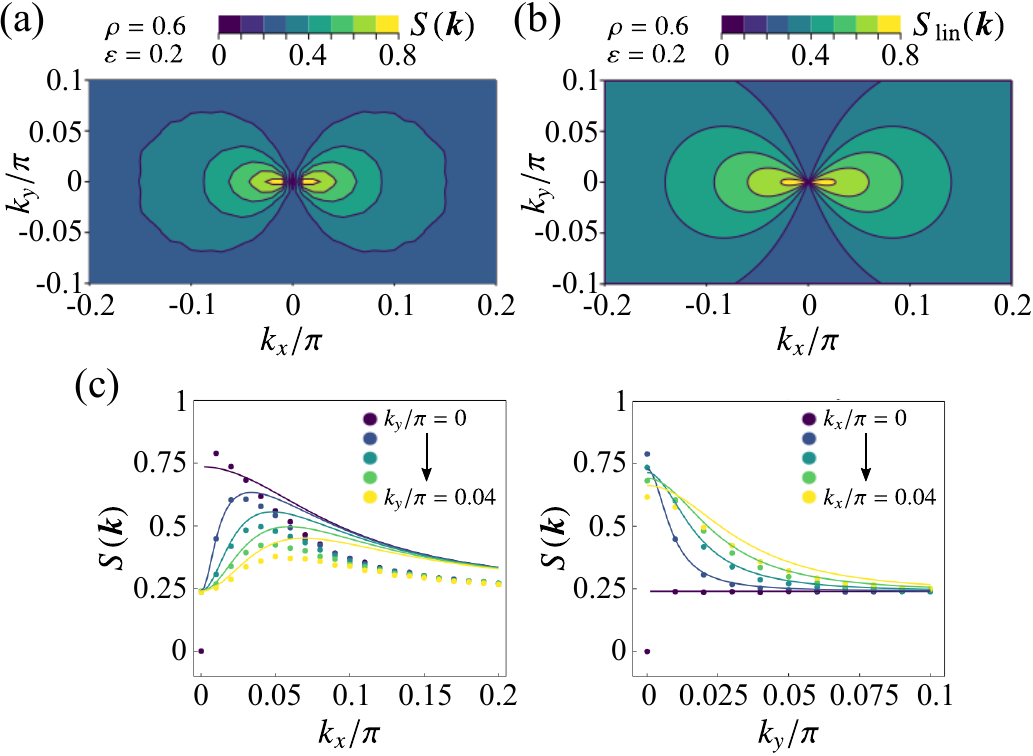}
\caption{(a) Contour plot of the structure factor $S(\bm{k})$ obtained numerically at $\rho = 0.6$ and $\varepsilon = 0.2$ (homogeneous state).
(b) Contour plot of the linearized structure factor $S_\mathrm{lin} (\bm{k})$ [Eq.~\eqref{Eq:LinStrFac}] for the same parameters as (a).
(c) Quantitative comparison between $S(\bm{k})$ (dots) and $S_\mathrm{lin}(\bm{k})$ (lines) for the same parameters as (a).
Note $S(\bm{k} = \bm{0}) = 0$ due to the particle number conservation.
For (a) and (c), we used $L_x = L_y = 200$.}
\label{Fig:CMIPS_homo}
\end{figure}

According to the studies on driven lattice gas models and coarse-grained Langevin models~\cite{Schmittmann1995,Schmittmann1998,Zia2010}, long-range density correlation is generically believed to appear in the non-equilibrium steady-state with spatial anisotropy of dynamics.
To examine whether the ALG shows long-range density correlation in the homogeneous steady-state, we calculate the structure factor $S (\bm{k}) := \sum_j \exp (- i \bm{k} \cdot \bm{r}_j) C(\bm{r}_j)$, where $C (\bm{r}) := (L_x L_y)^{-1} \sum_i \braket{[n (\bm{r}_i + \bm{r}) - \rho] [n (\bm{r}_i) - \rho]}$ is the correlation function that should be short-ranged in the equilibrium limit ($\varepsilon \to 0$).
As illustrated in Fig.~\ref{Fig:CMIPS_homo}(a) and colored dots in Fig.~\ref{Fig:CMIPS_homo}(c) for $L_x = L_y = 200$, $\rho = 0.6$, and $\varepsilon = 0.2$, we find a singularity of $S (\bm{k})$ at $\bm{k} = \bm{0}$, i.e., $S (k_x \to 0, k_y = 0) > S (k_x = 0, k_y \to 0)$, which means that the long-range density correlation exists as in driven lattice gas models~\cite{Schmittmann1995}.

To understand the singularity of $S (\bm{k})$, we apply the path-integral method~\cite{Lefevre2007,Martin1973,Janssen1976,Dominicis1976} and derive the Langevin equation for the spin-density field $\rho_{s} (\bm{r}, t)$ (see Appendix~\ref{App:Pathintegral}):
\begin{align}
\partial_t \rho_s = & J (\nabla^2 \rho_s - \rho_{-s} \nabla^2 \rho_s + \rho_s \nabla^2 \rho_{-s}) \nonumber \\
& - 2 s \varepsilon J \partial_x [(1 - \rho_+ - \rho_-) \rho_s] - h (\rho_s - \rho_{-s}) + \xi_s,
\label{Eq:CGLangevin}
\end{align}
where the lattice constant is set to unity, $\braket{\xi_s (\bm{r}, t)} = 0$, and $\braket{\xi_s (\bm{r}, t) \xi_{s'} (\bm{r}', t')} = \delta (t - t') M_{s, s'} \delta (\bm{r} - \bm{r}')$ with a differential operator $M_{s, s'} := \delta_{s, s'} [- 2 J \nabla \cdot (1 - \rho_+ - \rho_-) \rho_s \nabla] + (2 \delta_{s, s'} - 1) h (\rho_+ + \rho_-)$.
Linearizing Eq.~\eqref{Eq:CGLangevin}~\cite{Demery2014,Poncet2017,Poncet2021} and adiabatically eliminating the fast variable $\rho_+ (\bm{r}, t) - \rho_- (\bm{r}, t)$, we obtain the linear Langevin equation for the density fluctuation $\varphi (\bm{r}, t) := \rho_+ (\bm{r}, t) + \rho_- (\bm{r}, t) - \rho$, which can be solved in the Fourier space using $\varphi (\bm{k}, t) := \int d^2 \bm{r} \exp (- i \bm{k} \cdot \bm{r}) \varphi (\bm{r}, t)$ (see Appendix~\ref{App:Linearization}).
Within these approximations, we may calculate the structure factor $S_\mathrm{lin} (\bm{k}) := (L_x L_y)^{-1} \lim_{t \to \infty} \braket{|\varphi (\bm{k}, t)|^2}$, leading to
\begin{align}
& S_\mathrm{lin} (\bm{k}) = (1 - \rho) \rho \nonumber \\
& \times \frac{ [2 h + J (1 - \rho) \bm{k}^2] \bm{k}^2 + 4 \varepsilon^2 J (1 - \rho) {k_x}^2}{[2 h + J (1 - \rho) \bm{k}^2] \bm{k}^2 - 4 \varepsilon^2 J (1 - \rho) (2 \rho - 1) {k_x}^2}.
\label{Eq:LinStrFac}
\end{align}
As shown in Fig.~\ref{Fig:CMIPS_homo}(b) and colored lines in Fig.~\ref{Fig:CMIPS_homo}(c), $S_\mathrm{lin} (\bm{k})$ captures the qualitative feature observed in the simulation.
In particular, the singularity at $\bm{k} = \bm{0}$ is quantified~\cite{Schmittmann1998} by $S_\mathrm{lin} (k_x \to 0, k_y = 0) / S_\mathrm{lin} (k_x = 0, k_y \to 0) - 1 = 4 \varepsilon^2 J \rho (1 - \rho) / [h - 2 \varepsilon^2 J (1 - \rho) (2 \rho - 1)]$, which is nonzero if $\varepsilon \neq 0$.
Thus, the spatial anisotropy associated with the detailed balance violation in the ALG leads to the long-range density correlation.

\begin{figure*}[t]
\centering
\includegraphics[scale=0.8]{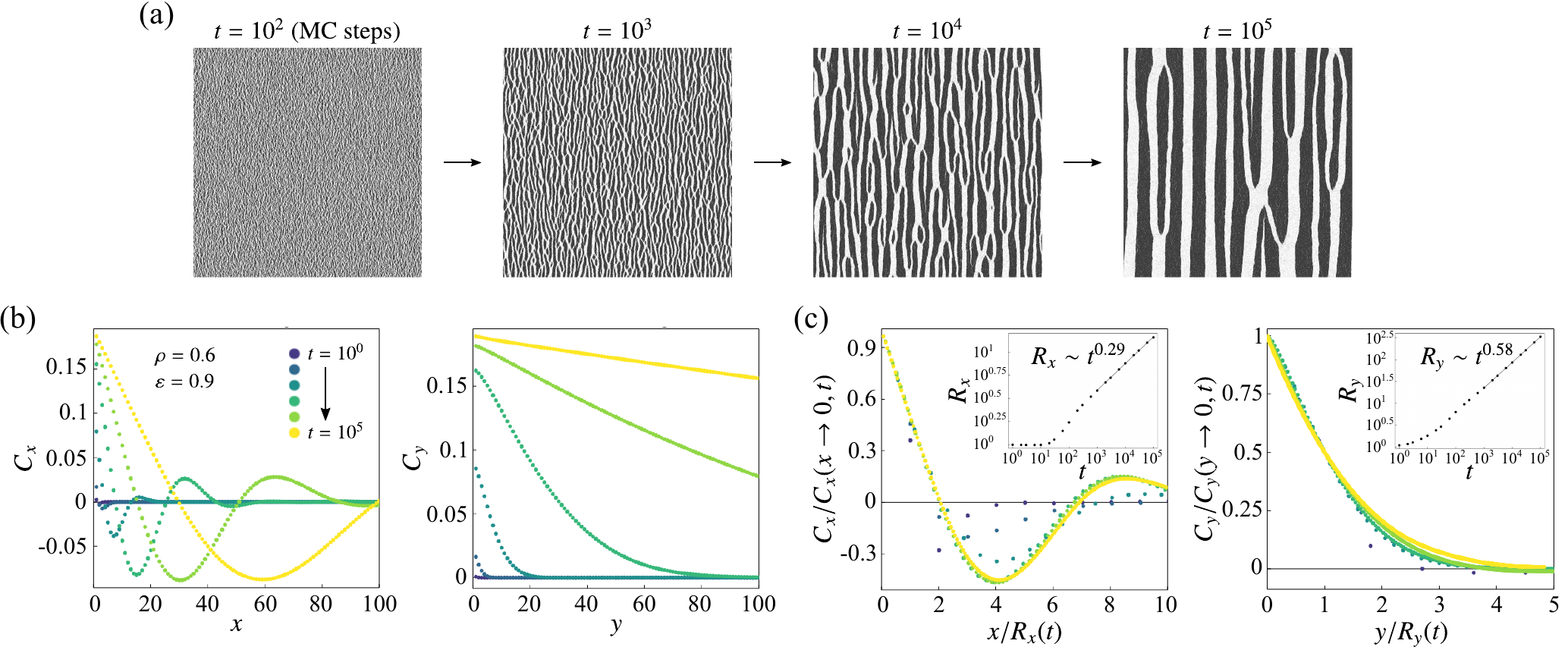}
\caption{(a) Coarsening process toward phase separation.
Each snapshot is a quarter square of the original system with $L_x = L_y = 3200$.
(b) Space/time-dependence of the density correlation functions along the $x$-axis ($C_x$) and the $y$-axis ($C_y$).
(c) Rescaled correlation functions as functions of the rescaled coordinates.
Time evolution of the domain size $R_{x (y)} (t)$ is shown in the inset, where the fitted line for the well-scaled region ($10^3 \ \mathrm{MC \ steps} \leq t \leq 10^5 \ \mathrm{MC \ steps}$) is also shown.
For all figures, we set $\rho = 0.6$ and $\varepsilon = 0.9$.}
\label{Fig:CMIPS_PS}
\end{figure*}

\begin{figure}[t]
\centering
\includegraphics[scale=0.8]{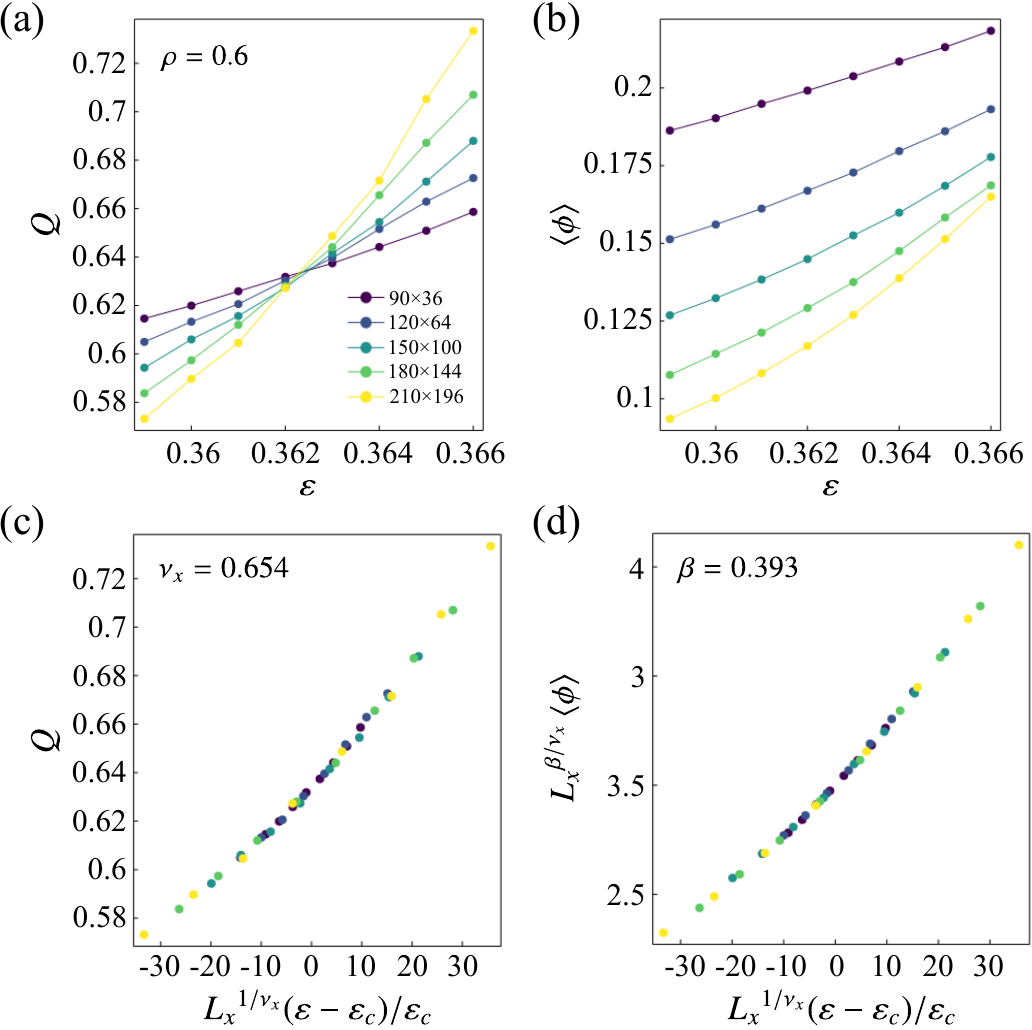}
\caption{(a) $\varepsilon$-dependence of the Binder ratio $Q$ for $\rho = 0.6$ and different system sizes with fixed $L_y / {L_x}^2 = 1 / 15^2$.
The solid lines are guides for the eyes.
(b) $\varepsilon$-dependence of $\braket{\phi}$ for the same parameters as (a).
(c) $Q$ as a function of the rescaled $\varepsilon$ with the best-fitted $\varepsilon_{c}$ $(\simeq 0.362)$ and $\nu_x$ $(\simeq 0.654)$.
(d) Rescaled $\braket{\phi}$ as a function of the rescaled $\varepsilon$ with the best-fitted $\beta$ $(\simeq 0.393)$ and the same values of $\varepsilon_{c}$ and $\nu_x$ as (c).}
\label{Fig:CMIPS_FSS}
\end{figure}

\subsection{Dynamic scaling in the phase-separated state}

We next investigate how the anisotropy appears in the dynamics of the PS state by focusing on the coarsening process toward phase separation [Fig.~\ref{Fig:CMIPS_PS}(a)].
We introduce $C_x (x, t) := C (x, y = 0, t)$ and $C_y (y, t) := C (x = 0, y, t)$, where $C (\bm{r}, t)$ is the time-dependent density correlation function.
Defining the typical domain size $R_x (t)$ along the $x$-axis as $C_x (R_x (t), t) = C_x (x \to 0, t) / 2$ and $R_y (t)$ in a similar way~\cite{Yeung1992}, we examine the rescaled correlation function $C_x (x, t) / C_x (x \to 0, t)$ as a function of $x / R_x (t)$ and the counterpart for $C_y (y, t)$.

For $L_x = L_y = 3200$, $\rho = 0.6$, and $\varepsilon = 0.9$, we find a good scaling behavior for $10^3 \lesssim t \lesssim 10^5$, where time $t$ is measured in units of 1 MC step [Figs.~\ref{Fig:CMIPS_PS}(b) and (c)].
Moreover, in the same time range, the growth dynamics shows an anisotropic power law as $R_x (t) \sim t^{\alpha_x}$ and $R_y (t) \sim t^{\alpha_y}$ with $\alpha_x < \alpha_y$ [insets in Fig.~\ref{Fig:CMIPS_PS}(c)].
Such anisotropic growth law with $\alpha_x < \alpha_y$ holds for different values of $\rho$ or $\varepsilon$ (see Appendix~\ref{App:Growth}).

\subsection{Critical point properties}
\label{Sec:CMIPS_FSS}

Recent simulations~\cite{Partridge2019,Maggi2021} and theories~\cite{Partridge2019} of the MIPS transition have suggested that the \textit{isotropic} MIPS critical point seems to show the Ising universality, i.e., the universality for equilibrium phase separation.
In contrast, effects of spatial anisotropy that we have described both in the homogeneous and PS states suggest that the universality of the \textit{anisotropic} MIPS critical point in the ALG is different from the Ising universality.
For the \textit{isotropic} MIPS, it is in fact still unclear whether the critical point generically belongs to the Ising universality class~\cite{Siebert2018,Dittrich2021}, since the macroscopic MIPS may be replaced by the microphase separation, or the bubbly phase separation~\cite{Tjhung2018,Caballero2018}, as observed in large-scale simulations~\cite{Shi2020}.
In our ALG, we did not find evidence of the anisotropic counterpart of the bubbly phase separation even in relatively large systems: $(L_x, L_y) = (1200, 400)$ (see Appendix~\ref{App:Growth}).

According to the studies on anisotropic nonequilibrium systems~\cite{Schmittmann1995}, there may exist two different exponents related to the divergence of the correlation length at criticality, $\nu_x$ and $\nu_y$, in two dimensions.
Based on Eq.~\eqref{Eq:CGLangevin}, we find that the effective model which should describe the critical dynamics of the ALG coincides with that of the randomly driven or two-temperature lattice gas model (see Appendix~\ref{App:Criticalmodel}), in which case the exponents satisfy $\nu_y / \nu_x \simeq 2$~\cite{Schmittmann1991,Schmittmann1993,Praestgaard1994,Praestgaard2000}.
To numerically estimate the critical exponents $\beta$, $\nu_x$, and $\nu_y$ for the ALG, we assume $\nu_y / \nu_x = 2$ and use the anisotropic finite-size scaling analysis~\cite{Leung1991,Wang1996}.
Briefly, we consider the scaling hypothesis as $\braket{\phi^n} = {L_x}^{- n \beta / \nu_x} F_n (L_x^{1 / \nu_x} (\varepsilon - \varepsilon_{c}), S)$, where $F_n$ is a scaling function, $\varepsilon_{c}$ is the critical point, and $S := L_y / {L_x}^{\nu_y / \nu_x} = L_y / {L_x}^2$.
We take $S = 1 / 15^2$ with varying $L_x$.
We set $\rho = 0.6$ as a rough estimate of the bottom point of the binodal curve based on Figs.~\ref{Fig:CMIPS}(b) and (c), and similar results are obtained if we take $\rho = 0.65$ (see Appendix~\ref{App:FSS}).

We find that the Binder ratio $Q(\varepsilon, L_x) := \braket{\phi^2}^2 / \braket{\phi^4}$ shows a crossing point [Fig.~\ref{Fig:CMIPS_FSS}(a)], which is consistent with the scaling hypothesis.
Fitting $Q(\varepsilon, L_x)$ with second-order polynomials (see Appendix~\ref{App:FSS}), we obtain $\varepsilon_{c} \simeq 0.36238(4)$ and $\nu_x \simeq 0.65(1)$, where the value in the bracket is the fitting error on the last significant figure.
Then, fitting $\braket{\phi} (\varepsilon, L_x)$ [Fig.~\ref{Fig:CMIPS_FSS}(b)] in a similar way, we find $\beta \simeq 0.3928(8)$.
By rescaling, we confirm that $Q$ and $\braket{\phi}$ respect the scaling function, consistent with the scaling hypothesis [Figs.~\ref{Fig:CMIPS_FSS}(c) and (d)].
Note that slight changes of $\nu_x$ and $\beta$ (e.g., $\nu_x = 0.6$ and $\beta = 0.35$) still give consistent scaling results for the system sizes used here (see Appendix~\ref{App:FSS}).
The obtained values of $\nu_x$ and $\beta$ are comparable to those of the two-temperature lattice gas model [$\nu_x \simeq 0.62(3)$ and $\beta \simeq 0.33(2)$]~\cite{Praestgaard2000} and two-loop renormalization group calculation of the corresponding effective Langevin model ($\nu_x \simeq 0.626$ and $\beta \simeq 0.315$)~\cite{Schmittmann1991,Schmittmann1993,Praestgaard2000}.
Although the accurate determination of the critical exponents is beyond the scope of our simulations, the ALG shows consistent results with the two-temperature lattice gas model within the tested regime.

\begin{figure}[t]
\centering
\includegraphics[scale=0.8]{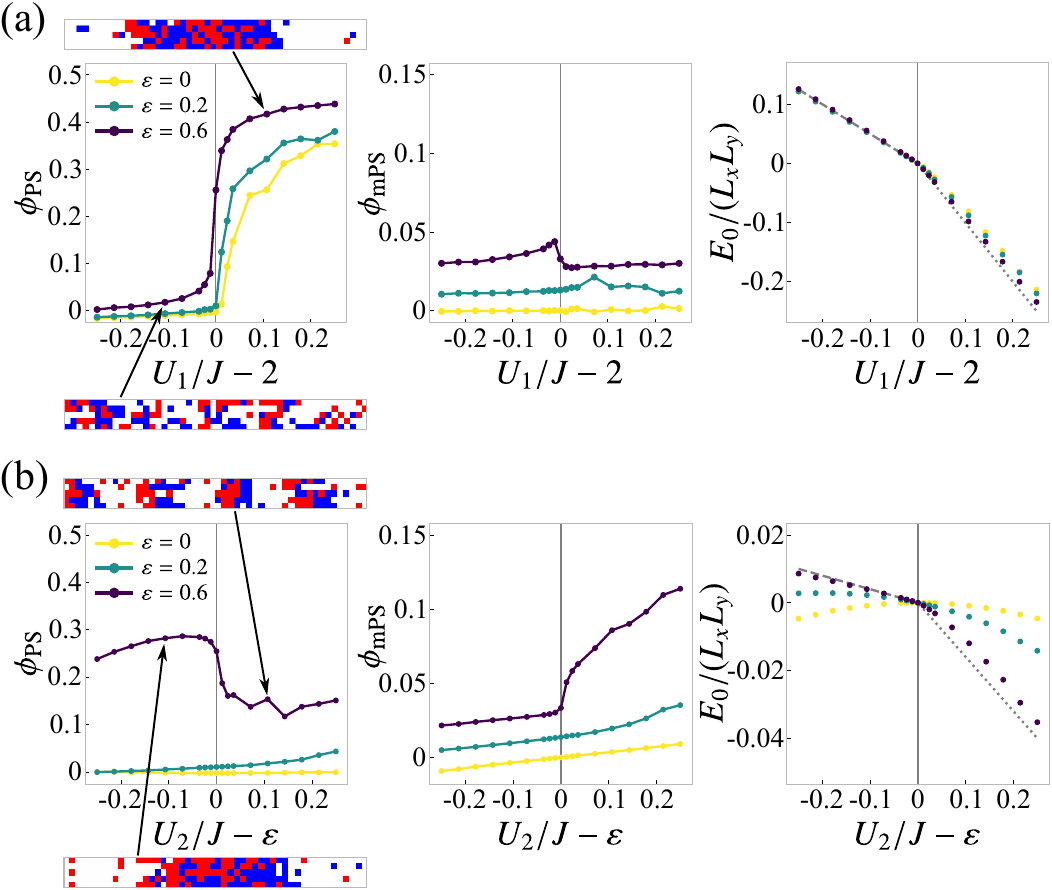}
\caption{
(a) $U_1$-dependence with $U_2 = \varepsilon J$ and (b) $U_2$-dependence with $U_1 = 2 J$ of the order parameters, $\phi_\mathrm{PS}$ and $\phi_\mathrm{mPS}$, and the ground-state energy, $E_0$, for $\rho = 0.5$, $h = 0.025 J$, and $\varepsilon = 0, 0.2, 0.6$ in $50 \times 5$ systems with typical configurations.
In the figures of $E_0$, we also plotted the analytical results of $\braket{H}_\mathrm{C}$ for (a) a disordered state (dashed) and a PS state (dotted) or (b) a mPS state with one (dashed) or four (dotted) clusters (see Appendix~\ref{Sec:Energy}).}
\label{Fig:QMIPS}
\end{figure}

\section{Quantum phase diagram and dynamical phase transition}

From the viewpoint of the full quantum model, the classical condition ($U_1 = 2J$ and $U_2 = \varepsilon J$) induces $E_0 = 0$.
The corresponding right eigenstate $\ket{\psi_0}$ is equivalent to the steady-state distribution of the \CModel{} (Fig.~\ref{Fig:Scheme}), and the left eigenstate is the coherent state, $\bra{\psi'_0} = \bra{P} := \bra{0} \exp (\sum_{i, s} a_{i, s})$.
For the case of $\varepsilon = 0$ and $U_2 = 0$, $H$ is Hermitian and equivalent to the ferromagnetic XXZ model with fixed magnetization~\cite{Matsubara1956} (see Appendix~\ref{Sec:corrXXZ}), where a first-order transition between the superfluid and phase-separated states occurs at the Heisenberg point ($U_1 = 2 J$)~\cite{Sariyer2019}.
The Heisenberg point is also special in that the right and left ground states are both coherent states.

To explore how the tendency toward MIPS comes into play beyond the classical condition, we conducted the diffusion Monte Carlo (DMC) simulation~\cite{Giardina2006} using elongated systems (e.g., $50 \times 5$).
In short, we run the Monte Carlo simulation for the \CModel{} but with the additional steps of re-sampling the states based on the calculated weights of the paths.
This works since the Hamiltonian can be divided into two parts $H = - W - D$, where $W := - H(J,\varepsilon,U_1=2J,U_2=\varepsilon J,h)$ corresponds to the classical dynamics and $D$, being a diagonal matrix, can be interpreted as the re-sampling weights (see Appendix~\ref{Sec:DMC}).
To discuss the phases, we focus on physical quantities which are functions of the configuration of the particles, $A (\{ \hat{n}_{i, s} \})$, and calculate $\braket{A}_\mathrm{C} := \braket{P | A (\{ \hat{n}_{i, s} \}) | \psi_0} / \braket{P | \psi_0}$.
PS states are characterized by
\begin{eqnarray}
\phi_\mathrm{PS} := (L_x L_y)^{-1} \sum_{\langle i, j \rangle} \braket{(\hat{n}_i - \rho) (\hat{n}_j - \rho)}_\mathrm{C}.
\end{eqnarray}
For microphase-separated (mPS) states, in which the number of clusters is $O(L_x)$ [see the upper configuration in Fig.~\ref{Fig:QMIPS}(b)], we utilize $\phi_\mathrm{mPS}$ as the order parameter, which is the density of clusters with oppositely polarized edges: 
\begin{eqnarray}
\phi_\mathrm{mPS} := {L_x}^{-1} \sum_{i=1}^{L_x} \braket{\hat{m}_{i}^X (\hat{n}_{i + 1}^X - \hat{n}_{i - 1}^X)}_\mathrm{C},
\end{eqnarray}
where $\hat{n}_{i}^X := {L_y}^{-1} \sum_{j=1}^{L_y} \hat{n}_{i \hat{x} + j \hat{y}}$ and $\hat{m}_{i}^X := {L_y}^{-1} \sum_{j=1}^{L_y} \hat{m}_{i \hat{x} + j \hat{y}}$.
In the large-size limit ($L_x, L_y \to \infty$), $\phi_\mathrm{PS} > 0$ and $\phi_\mathrm{mPS} = 0$ for the PS state, while $\phi_\mathrm{PS} > 0$ and $|\phi_\mathrm{mPS}| > 0$ for the mPS state.

\begin{figure}[t]
\centering
\includegraphics[scale=0.8]{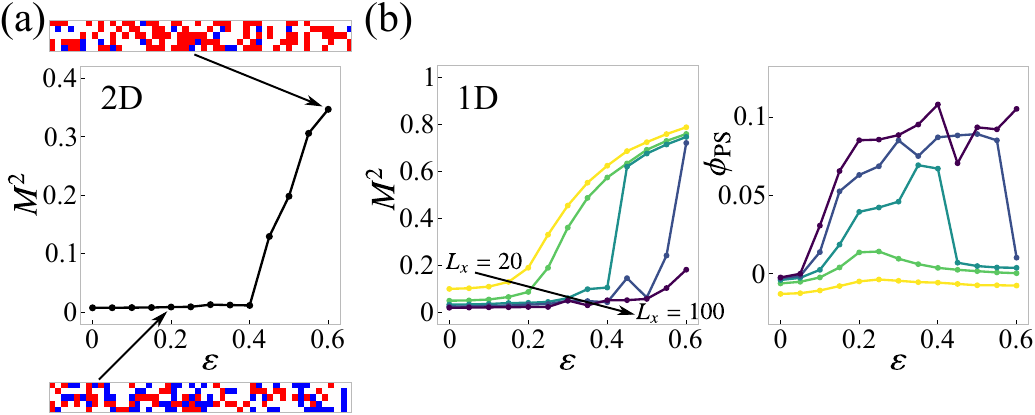}
\caption{
(a) $\varepsilon$-dependence of the squared magnetization $M^2$ and typical configurations in $50 \times 5$ systems.
(b) $\varepsilon$-dependence of $M^2$ and $\phi_\mathrm{PS}$ in 1D systems with $L_x = 20, 40, 60, 80, 100$.
In both (a) and (b), we set $\rho = 0.5$, $h = 0.025 J$, $U_1 = 2 J$, and $U_2 = 0$.}
\label{Fig:ep}
\end{figure}

\begin{figure}[t]
\centering
\includegraphics[scale=0.8]{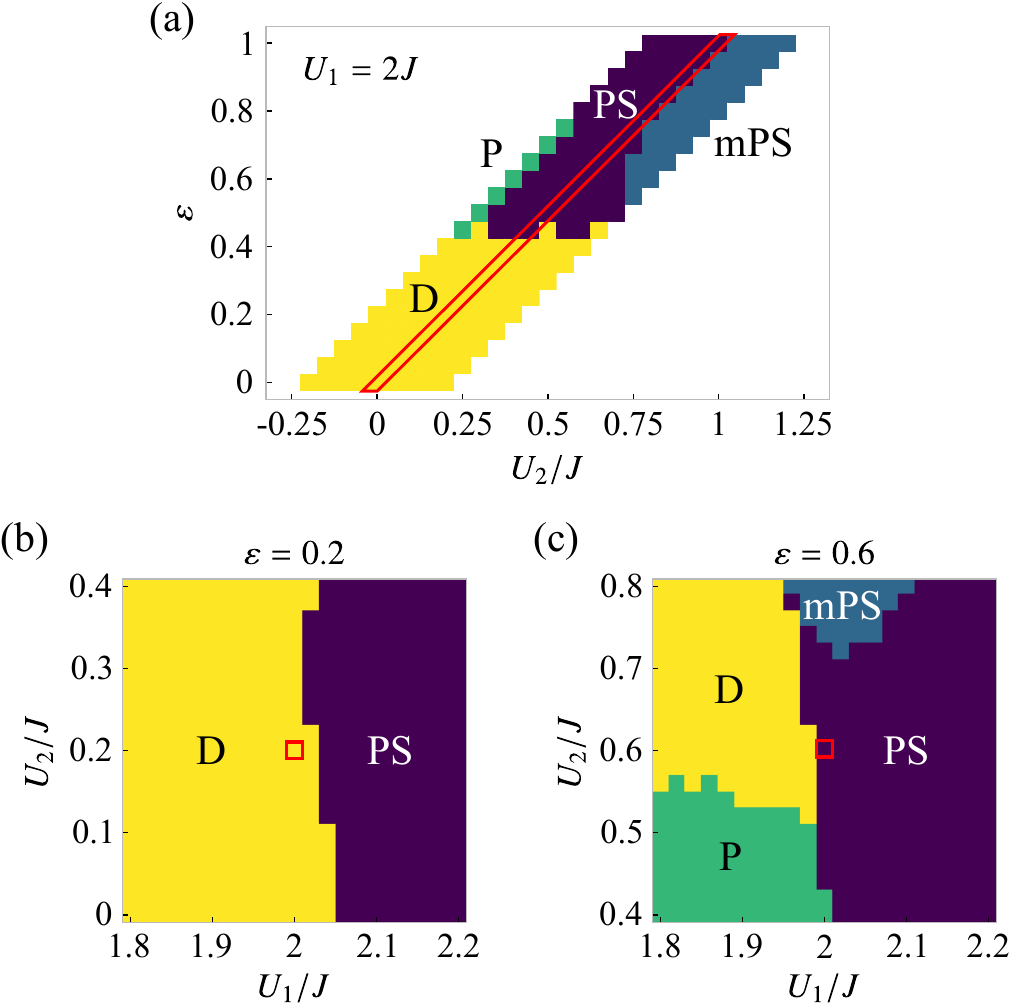}
\caption{Ground-state phase diagrams of the quantum model.
(a) $U_2$-$\varepsilon$ phase diagram for $U_1 = 2 J$ around the classical line (red box), with PS ($\phi_\mathrm{PS} > 0.1$ and $\phi_\mathrm{mPS} \leq 0.1$), mPS ($\phi_\mathrm{mPS} > 0.1$), P (polar, $M^2 > 0.1$), and D (disordered, otherwise) states.
(b) and (c) $U_1$-$U_2$ phase diagrams for $\varepsilon = 0.2$ and $0.6$, respectively, around the cross section of the classical line (red box).
In all figures, we set $\rho = 0.5$ and $h = 0.025 J$.}
\label{Fig:PD}
\end{figure}

\subsection{Quantum phase transitions}

We first find that there is a discontinuous phase transition induced by slightly increasing  $U_1$ from $2J$.
As shown in Fig.~\ref{Fig:QMIPS}(a), $\phi_\mathrm{PS}$ increases rapidly as a function of $U_1$ at around $U_1 = 2 J$ for a broad range of $\varepsilon$ $(= 0, 0.2, 0.6)$ and $U_2$ $(= \varepsilon J)$, with the ground-state energy $E_0$ having a kink at $U_1 = 2 J$.
This line of phase separation transition extends from the first-order transition in the XXZ model ($\varepsilon = 0$)~\cite{Matsubara1956,Sariyer2019}.
Second, for high enough $\varepsilon$ $(= 0.6)$, a drop in $\phi_\mathrm{PS}$ and an increase in $\phi_\mathrm{mPS}$ occur simultaneously as $U_2$ crosses $\varepsilon J$ [Fig.~\ref{Fig:QMIPS}(b)].
As also indicated from the typical configuration and the kink in $E_0$ [Fig.~\ref{Fig:QMIPS}(b)], this is expected to be a discontinuous transition between the PS and mPS states.
For low $\varepsilon$ $(= 0, 0.2)$, in contrast, we do not see this transition [Fig.~\ref{Fig:QMIPS}(b)].
We observed similar transitions in a one-dimensional (1D) setup, even though the corresponding classical model does not show MIPS (see Appendix~\ref{Sec:1Dmodel}).

Next, we consider increasing $\varepsilon$ while fixing $U_1 = 2 J$ and $U_2 = 0$.
Intriguingly, we find that a ferromagnetic order appears without phase separation for high $\varepsilon$ $(\gtrsim 0.4)$, indicated by $M^2 := N^{-2} \braket{\left( \sum_i \hat{m}_i \right)^2}_\mathrm{C}$ [Fig.~\ref{Fig:ep}(a)].
Such polar order, which should be accompanied by flow due to the asymmetric hopping, is reminiscent of the flocking of self-propelled particles observed, e.g., in the Vicsek model~\cite{Vicsek1995}, although our model \eqref{Eq:Hamiltonian} does not include explicit polar interactions.

To investigate whether the polar order remains in larger systems, we further performed simulations in 1D systems.
The size-dependence of $M^2$ and $\phi_\mathrm{PS}$ in 1D systems [Fig.~\ref{Fig:ep}(b)] shows that the polar state is destabilized and instead the PS state appears as the system size becomes larger.
In addition, the discontinuous changes in $M^2$ and $\phi_\mathrm{PS}$ indicate bistability of the polar and PS states in finite systems.
Similarly, in large two-dimensional systems, the PS state can replace the polar state, as observed in the $U_2$-dependence of $M^2$ and $\phi_\mathrm{PS}$ for the system with size $50 \times 5$ (see Appendix~\ref{Sec:convergence}).
Therefore, we find that the non-Hermitian asymmetric hopping terms alone (with $U_2=0$) will lead to either the polar state or the PS state, which are the quantum analogs of the flocking and MIPS states, respectively.

In Fig.~\ref{Fig:PD}, we show the phase diagram for a system with size $30 \times 3$.
First, Fig.~\ref{Fig:PD}(a) is the $U_2$-$\varepsilon$ phase diagram around the classical line ($U_1 = 2 J$ and $U_2 = \varepsilon J$) indicated in red.
In addition to the classical MIPS, the PS-mPS transition occurs when crossing the classical line at high $\varepsilon$ [see Fig.~\ref{Fig:QMIPS}(b)].
Next, Figs.~\ref{Fig:PD}(b) and (c) display the $U_1$-$U_2$ phase diagrams around the classical line.
For low $\varepsilon$ $(= 0.2)$ [Fig.~\ref{Fig:PD}(b)], we find that the $U_1$-induced phase separation transition [Fig.~\ref{Fig:QMIPS}(a)] occurs robustly against $U_2$-perturbation from the classical line.
In contrast, for high $\varepsilon$ $(= 0.6)$ [Fig.~\ref{Fig:PD}(c)], slight changes in $U_1$ and $U_2$ around the classical line can lead to the mPS and polar states.

\begin{figure}[t]
\centering
\includegraphics[scale=0.8]{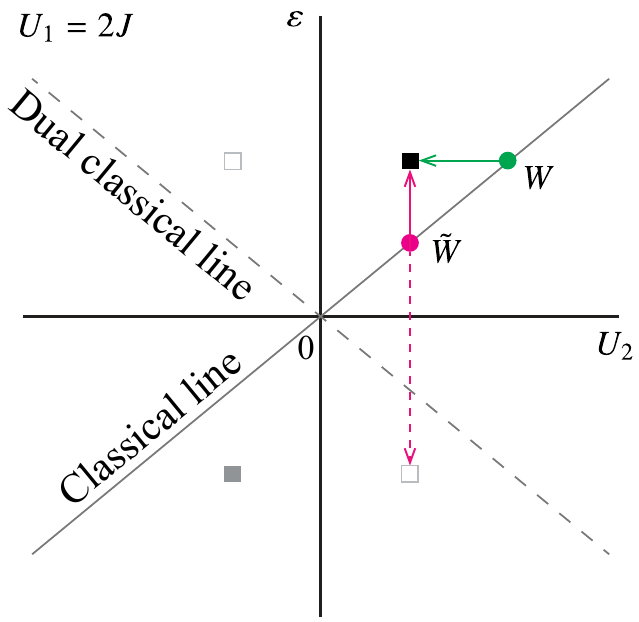}
\caption{Schematic of the $U_2$-$\varepsilon$ plane at $U_1 = 2J$. The Hamiltonian has two symmetries (see the main text), meaning that the points indicated by squares all have the same value of $E_0$. The  classical line ($U_2 = \varepsilon J$) and the dual classical line ($U_2 = -\varepsilon J$) have $E_0 = 0$.
The same Hamiltonian (e.g., black square) can be described in multiple ways of classical stochastic dynamics (e.g., $W$ and $\tilde{W}$) with bias (e.g., green and magenta arrows).}
\label{Fig:cline}
\end{figure}

The DMC simulation becomes less reliable for the parameter regions far away from the classical line.
Nevertheless, there are symmetries in this system that indicate the positions of the phase boundaries in a wider parameter region (Fig.~\ref{Fig:cline}).
First, we have $E_0 (J,-\varepsilon,U_1, -U_2,h) = E_0 (J,\varepsilon,U_1, U_2,h)$ which is due to $H (J,\varepsilon,U_1, U_2,h) = \hat{U}^\dag H (J,-\varepsilon,U_1, -U_2,h) \hat{U}$, where $\hat{U}$ is the unitary operator of spin reversal.
We also have $E_0 (J,-\varepsilon,U_1, U_2,h) = E_0 (J,\varepsilon,U_1, U_2,h)$ since $H (J,-\varepsilon,U_1, U_2,h)^\dag = H (J,\varepsilon,U_1,U_2,h)$.
Since the analytical property of $E_0$ indicates the positions of the phase boundaries, we expect that the boundaries calculated in Fig.~\ref{Fig:PD} may have corresponding phase boundaries in $\varepsilon<0$ and/or $U_2<0$ regions.
For example, there should be a transition for large enough $|\varepsilon|$ in crossing the dual classical line defined by $U_1 = 2 J$ and $U_2 = -\varepsilon J$, which is where $E_0=0$ and $\ket{\psi_0}=\ket{P}$ (Fig.~\ref{Fig:cline}).

\subsection{Connection to dynamical phase transition in classical kinetics}

The scheme of the DMC implies an interesting connection between the quantum model and the classical kinetics.
For the \CModel{} with the transition rate matrix $W$, we denote the configuration of the particles at time $t$ as $\mathcal{C}_t = \{ n_{i,s} (t) \}$, and its stochastic trajectory as $\mathcal{C}_t = \mathcal{C}_k \ (t_k \leq t < t_{k+1})$ with $t_k$ being the time point of the $k$-th jump.
For a path-dependent quantity $\bar{B} _\tau := \int_0 ^\tau dt B_{\mathcal{C}_t,\mathcal{C}_t} + \sum_k B_{\mathcal{C}_k,\mathcal{C}_{k+1}}$ defined using an arbitrary real matrix $B$ that acts on the Fock space, we introduce
\begin{eqnarray}
    \lambda^W(B) := \lim_{\tau \to \infty}  \frac{1}{\tau} \ln \braket{ \exp ( \bar{B}_\tau ) } ^W,  \label{Eq:gscgf}
\end{eqnarray}
where the ensemble average $\braket{ \cdots }^W$ is taken over the trajectories in the \CModel{}.
$\lambda^W(B)$ is equivalent to the dominant eigenvalue of a biased transition rate matrix~\cite{lebowitz1999,garrahan2007}:
\begin{eqnarray}
W^{B}_{\mathcal{C},\mathcal{C}'} := (1- \delta_{\mathcal{C},\mathcal{C}'} ) W_{\mathcal{C},\mathcal{C}'} e^{B_{\mathcal{C},\mathcal{C}'}} + \delta_{\mathcal{C},\mathcal{C}'} (W_{\mathcal{C},\mathcal{C}'} + B_{\mathcal{C},\mathcal{C}'}) \end{eqnarray}
Typical paths that appear in the biased dynamics can become dramatically different from the original dynamics, which is the hallmark of dynamical phase transition that can be captured by the (non-)analytical behavior of $\lambda^W(B)$~\cite{garrahan2007}.
Biased kinetics and dynamical phase transition have been studied with interests in exploring glassy systems and in characterizing phases in models of active matter~\cite{Whitelam2018,Nemoto2019,Tociu2019}.

The quantum Hamiltonian \eqref{Eq:Hamiltonian} can be interpreted as the transition rate matrix with bias by writing $H=-W^B$, where the bias is $B= u_1 F + u_2 G$ with  $F_{ \mathcal{C},\mathcal{C}'} := \bra{\mathcal{C}}\sum_{\langle i,j \rangle } \hat{n}_i \hat{n}_j \ket{\mathcal{C'}}$ and $G_{\mathcal{C},\mathcal{C}'} := \bra{\mathcal{C}}\sum_{i} \hat{m}_i (\hat{n}_{i+\hat{x}} - \hat{n}_{i-\hat{x}}) \ket{\mathcal{C'}}$ being diagonal matrices.
Here, $\ket{\mathcal{C}}$ is the Fock-space basis corresponding to the configuration $\mathcal{C}$, and $u_1 := U_1-2J, u_2 := U_2-\varepsilon J$ quantifies the displacement from the classical line.
We then arrive at
\begin{align} 
 E_0(J,\varepsilon,U_1,U_2,h) = -\lambda^{W}( u_1 F + u_2 G),
\label{Eq:EnergySCGF}
\end{align}
which means that the quantum phase transitions, captured by the property of $E_0$, are equivalent to the dynamical phase transitions induced by the bias $u_1 F + u_2 G$.
The bias here has a clear interpretation: increasing $u_1$ and $u_2$ favors larger $\phi_\mathrm{PS}$ and $\phi_\mathrm{mPS}$, respectively.

More generally, we may consider an arbitrary pair of a transition rate matrix $\tilde{W}$ and bias $\tilde{B}$ that satisfies $H = -\tilde{W}^{\tilde{B}}$.
One interesting choice is $\tilde{W} = -H(J = U_1/2,\varepsilon=2 U_2/ U_1,U_1,U_2,h)$, which is a matrix with the same diagonal elements as $-H$ but with the off-diagonal elements tuned so that $\sum_\mathcal{C} \tilde{W}_{\mathcal{C},\mathcal{C}'} =0$. The corresponding bias will be 
\begin{eqnarray}
    \tilde{B}_{\mathcal{C},\mathcal{C}'}=|V_{\mathcal{C},\mathcal{C}'}| \ln \frac{J}{J_0} +  \ln \frac{1+\varepsilon V_{\mathcal{C},\mathcal{C}'}}{1+\varepsilon_0 V_{\mathcal{C},\mathcal{C}'}},
\end{eqnarray}
which is non-diagonal and non-Hermitian (Fig.~\ref{Fig:cline}).
Here, $V$ is a skew-Hermitian matrix given by
\begin{eqnarray}
    V_{\mathcal{C},\mathcal{C}'} =  \sum_{i, s} s \bra{\mathcal{C}}  ( a_{i, s}^\dag a_{i - \hat{x}, s} - a_{i, s}^\dag a_{i + \hat{x}, s} ) \ket{\mathcal{C}'}
\end{eqnarray}
Introducing the entropy production by its commonly used definition~\cite{lebowitz1999}:
\begin{eqnarray}
    \sigma_{\mathcal{C},\mathcal{C}'} (W') :=  \ln \frac{ W'_{\mathcal{C},\mathcal{C}'}}{W'_{\mathcal{C}',\mathcal{C}}},
\end{eqnarray}
we find,
\begin{eqnarray}
    \tilde{B} - \tilde{B}^\dag =\sigma (\tilde{W}^{\tilde{B}} )- \sigma(\tilde{W}),
\end{eqnarray}
which indicates that the difference of entropy production defined in the biased and unbiased kinetics is exactly the non-Hermiticity of the bias $\tilde{B}$.
We also note that there is a fluctuation theorem-like relation~\cite{lebowitz1999}:
\begin{eqnarray}
    \lambda^{\tilde{W}} (\tilde{B})=\lambda^{\tilde{W}} (\tilde{B}^\dag -\sigma(\tilde{W})),
\end{eqnarray}
which follows from $( \tilde{W}^{\tilde{B}} )^\dagger = \tilde{W} ^{-\sigma(\tilde{W}) + \tilde{B}^\dagger}$.
This symmetry, which is nothing but the $E_0 (J,-\varepsilon,U_1, U_2,h) = E_0 (J,\varepsilon,U_1, U_2,h)$ symmetry, is depicted as magenta arrows in Fig.~\ref{Fig:cline}.

The $\varepsilon$-dependent transition toward the flocking phase (Fig.~\ref{Fig:ep}) can be understood as the consequence of biasing the kinetics toward larger $\tilde{B}$, which encourages more spin-dependent asymmetric hopping and therefore dissipation.
Consistent with this, dynamical phase transition induced by biasing toward higher dissipation has been reported in the studies of active Brownian particles~\cite{Whitelam2018,Tociu2019,Nemoto2019}.

\section{Relevance to experiments}

Lastly, we describe an example procedure to implement the model~\eqref{Eq:Hamiltonian} and observe the activity-induced phase transitions in a quantum experiment.
\subsection{Implementation of the quantum model}

The basic model is a two-component Bose-Hubbard model on a square lattice, which is realized with bosonic ultracold atoms (e.g., $^{87}$Rb) in optical lattices~\cite{Gross2017}.
The two components can be realized as the two internal states of atoms, and the Hubbard interaction is controllable via the Feshbach resonance. We also require strong repulsive interaction to reach the hard-core limit. 
Other ingredients to be implemented are the transverse magnetic field, nearest-neighbor interaction, and spin-dependent asymmetric hopping. 

The transverse magnetic field can be implemented by a coherent coupling between the two internal states. Such coherent coupling is well-studied and widely used in two-component bosonic atomic gases~\cite{Gross2017}. 
For the nearest-neighbor interactions, although they are generally more difficult to implement in optical lattice systems compared with on-site interactions, there have been various proposals such as the use of optical cavity~\cite{Landig2016}, Rydberg states~\cite{Browaeys2020}, dipolar interaction~\cite{Trefzger2011}, and Floquet engineering~\cite{Zhao2019} to overcome the difficulty.
Another idea to realize the attractive interaction under the hard-core condition using dissipation is to consider an attractive Bose-Hubbard model with strong two-body loss, which can be introduced both intrinsically~\cite{Syassen2008} and artificially, e.g., via photoassociation~\cite{Tomita2017}.
In the Zeno limit, the quantum Zeno effect suppresses the double occupancy and the hard-core condition should be effectively satisfied. A similar phenomenon in the case of three-body loss has been observed in experiment~\cite{Mark2012}. 

The spin-dependent asymmetric hopping is qualitatively different from the other terms since it is non-Hermitian. However, as discussed in Ref.~\cite{Gong2018}, it is possible to implement the non-Hermitian effect by using the dissipative optical lattice setup. 
The typical description of dissipative cold atomic systems is by a Lindblad-type quantum master equation~\cite{Daley2014}, which is $d \rho(t)/dt=-i [H_0, \rho(t)] + \sum_k \mathcal{D}[L_k]\rho(t)$, where $\mathcal{D}[L]\rho(t)=L \rho(t) L^\dagger - \{L^\dagger L, \rho(t) \}/2$.
Under postselection, where we only focus on the experimental data that the loss process did not happen, this master equation is simplified as $d\rho(t)/dt=-i (H_\eff \rho(t)-\rho(t)H_\eff^\dagger)$.
Here, the effective Hamiltonian is defined as
\begin{align}
    H_\eff=H_0 - \frac{i}{2} \sum_{k} L_k^\dagger L_k. \label{eq:Heff}
\end{align}
Intuitively, this effective Hamiltonian contains the back action of the dissipation due to the restriction of the Hilbert space to the situation where the loss event did not happen, which becomes the origin of non-Hermiticity. This non-Hermitian Hamiltonian has been well-examined in the studies of the quantum trajectory method, which is an efficient approach to simulate the dynamics of open quantum systems~\cite{Dalibard1992,Dum1992,Carmichael_book,Daley2014}.
The form of the effective Hamiltonian \eqref{eq:Heff} suggests that we can engineer the non-Hermitian Hamiltonian by choosing the adequate dissipators $\{ L_k \}$. To engineer the spin-dependent asymmetric hopping, we can use the following dissipators:
\begin{align}
    L_{k} = \sqrt{2 \varepsilon J} (a_{j, s} + i s a_{j+\hat{x}, s}), \label{eq:diss}
\end{align}
where $k$ denotes a pair of the indices $(j, s)$. Assuming these dissipators, we obtain 
\begin{align}
    - \frac{i}{2} \sum_{k} L_k^\dagger L_k= - \varepsilon J \sum_{j, s} s ( a_{j, s}^\dag a_{j - \hat{x}, s} - a_{j, s}^\dag a_{j + \hat{x}, s} ) - 2i \varepsilon JN
\end{align} 
where $N$ denotes the total number operator $N= \sum_{j, s} \hat{n}_{j, s}$. The first term is nothing but the spin-dependent asymmetric hopping in the model~\eqref{Eq:Hamiltonian}. 
Since we consider a subspace with a fixed total particle number, the second term only gives a constant energy shift. 

\begin{figure}
\centering
\includegraphics[scale=0.25]{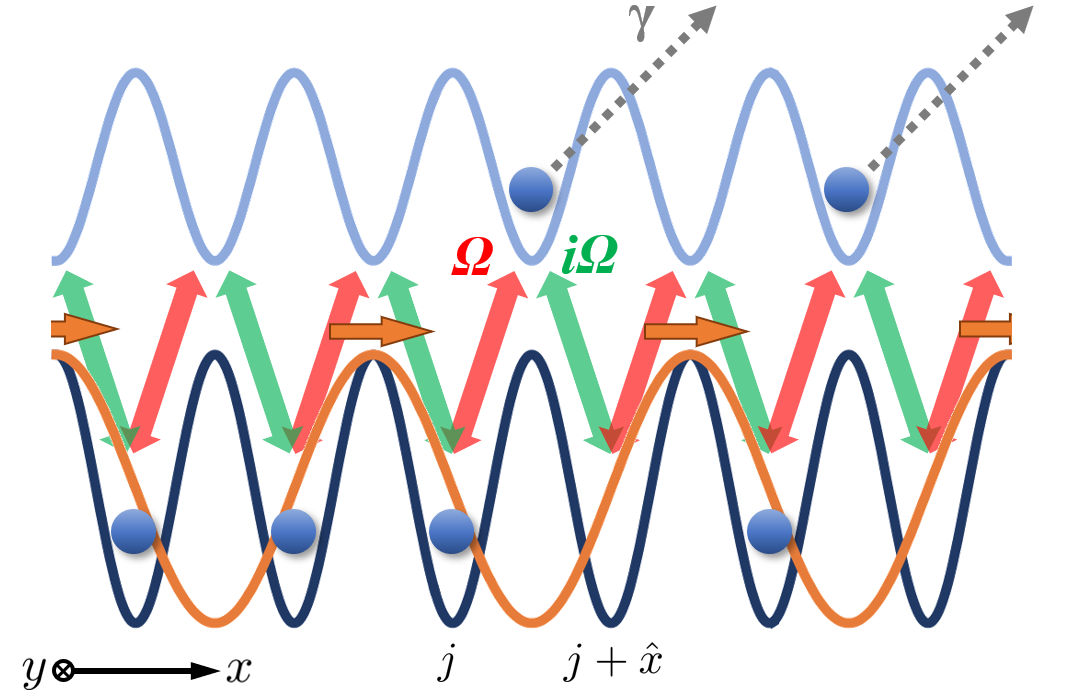}
\caption{Experimental implementation of the asymmetric hopping in cold atomic systems. The original optical lattice (dark blue), the dissipative optical lattice (light blue), the coherent coupling between two lattices (red and green arrows) and the running wave in the $x$-direction (orange line) are introduced and bosonic atoms (blue balls) are loaded in the optical lattices.}
\label{Fig:optical_lattice}
\end{figure}

Lastly, we note on how to implement the dissipator~\eqref{eq:diss} in experiments.
The basic idea is to introduce a nonlocal coherent coupling to an auxiliary dissipative lattice, schematically shown in Fig.~\ref{Fig:optical_lattice}~\cite{Gong2018}.
The coherent coupling to the dissipative lattice displaced by half of the lattice constant
naturally induces the hopping from the $j$- and $(j+\hat{x})$-sites, which becomes the origin of the nonlocal loss. This setting of half-lattice is possible by using the internal atomic states with opposite Stark shifts. For instance, $^1$S$_0$ and $^3$P$_0$ of $^{174}$Yb atoms have the opposite Stark shift~\cite{Gong2018}. Writing down the master equation within the tight-binding approximation and eliminating the fast decay mode, we can obtain the nonlocal one-body loss~\cite{Gong2018}. In addition, we need to introduce a running wave laser whose wavelength is equal to that of the lattice constant. This running wave provides the phase difference between the couplings at the $j$- and $(j+\hat{x})$-sites. 
Taking this effect into account, we obtain \eqref{eq:diss} except for the spin-dependency. 
To implement the spin-dependent asymmetric hopping, we further require oppositely-directed running wave lasers coupled to each spin component, as in the case of spin-selective optical lattice~\cite{Gross2017}.

\subsection{Preparation of the ground state}

In this study, we discussed the properties of the ground state $\ket{\psi_0}$, which the eigenstate with the smallest real part of the energy eigenvalue.
In non-Hermitian systems, however, this state is not realized at low temperature since the state with the largest imaginary part of the energy eigenvalue will dominate in the long-time limit.

A workaround to this problem is to take the approach of adiabatic preparation. First, we prepare the Hermitian system ($\varepsilon=0$) and realize the low-temperature state via thermalization in a closed quantum system. Then, we introduce dissipation adiabatically, i.e., turn on the asymmetric hopping term very slowly. Thanks to the Perron-Frobenius theorem, the uniqueness and the realness of the ground state energy is guaranteed, and thus the energy gap $\Delta = |E_1-E_0|$ should remain non-zero in a finite system through this process at least for a finite time. 
Although the adiabatic theorem is invalid in the strict sense, it has been shown that, when there is a finite gap $\Delta$, the state keeps sitting on the same state for a finite time under varying the parameters slowly~\cite{Wang2018}.
We remark that a similar approach has also been used in previous works on non-Hermitian quantum many-body systems~\cite{Ashida2017, Yamamoto2019}.

\subsection{Measurable quantities and their relation to the results from the Monte Carlo simulation}
The most promising method to detect activity-induced phase transition such as MIPS is a quantum gas microscope (QGM)~\cite{Gross2017}. This enables us to measure the observable in a spatially resolved way. Using the observed quantities, we can calculate the order parameters of each phase transition. For instance, the indicator of MIPS, $\phi_\mathrm{PS}$, is calculated from the local density data. The technique of QGM is growing rapidly and the measurements in the Bose-Hubbard systems have already been conducted~\cite{Bakr2010}.

In real experiments in open quantum systems, the measurable quantity is $\braket{\cdots}_\mathrm{Q} := \braket{\psi_0 |  \cdots | \psi_0}$ rather than $\braket{\cdots}_\mathrm{C}$~\cite{Dalibard1992,Dum1992,Carmichael_book,Daley2014}.
Furthermore, typical cold atom experiments are in open boundary condition (OBC)~\cite{Gross2017}, in which case the exact mapping to a classical system does not exist (see Appendix~\ref{Sec:genq}).
To address these points, we conducted exact diagonalization for a small 1D system to check how the redefining the order parameters using $\braket{\cdots}_\mathrm{Q}$ and the different boundary conditions will change the result. For each of $\braket{\cdots}_\mathrm{Q}$ and $\braket{\cdots}_\mathrm{C}$, we define the order parameters, $\phi_\mathrm{PS}$, $\phi_\mathrm{mPS}$, and $M^2$. For $\braket{\cdots}_\mathrm{Q}$, we also define the order parameter for the superfluid (SF) state, which is characterized by the off-diagonal long-range correlation, as $\phi_\mathrm{SF} := {L_x}^{-1} \sum_s \sum_{|i-j|=L_x/2} \braket{ a_{i,s}^\dag a_{j,s} }_\mathrm{Q}$. Note that $\phi_\mathrm{SF}$ for $\braket{\cdots}_\mathrm{C}$ is meaningless since the SF order and the density order are equivalent ($\braket{a_{i,s}^\dag a_{j,s}}_\mathrm{C} = \braket{\hat{n}_{i,s} \hat{n}_{j,s}}_\mathrm{C}$). 

\begin{figure}[t]
\centering
\includegraphics[scale=0.8]{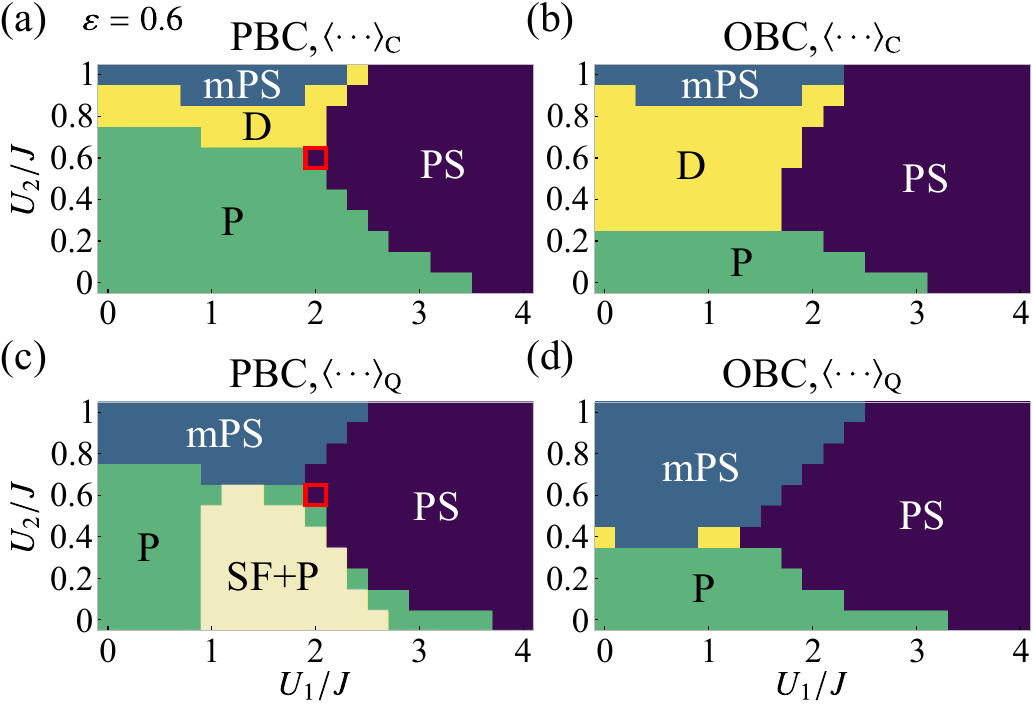}
\caption{$U_1$-$U_2$ phase diagrams for $\varepsilon = 0.6$ in small 1D systems ($L_x = 12$), with PS ($\phi_\mathrm{PS} > 0.05$ and $\phi_\mathrm{mPS} \leq 0.3$), mPS ($\phi_\mathrm{mPS} > 0.3$), P ($M^2 > 0.2$), SF (superfluid, $\phi_\mathrm{SF} > 0.2$), and D (otherwise) states.
The order parameters are calculated by exact diagonalization, using $\braket{\cdots}_\mathrm{C}$ [(a) and (b)] or $\braket{\cdots}_\mathrm{Q}$ [(c) and (d)], for the PBC [(a) and (c)] or OBC [(b) and (d)].
Superfluid states cannot be identified in DMC calculations or by using $\braket{\cdots}_\mathrm{C}$ (see the main text and Appendix~\ref{Sec:ED}).
In all figures, we set $\rho = 0.5$ and $h = 0.025 J$.}
\label{Fig:ED_main}
\end{figure}

As shown in the phase diagram~(Fig.~\ref{Fig:ED_main}), we found that all of the phases exist in the various setups, with an additional polar-SF phase which can be captured by an off-diagonal order parameter, indicating that experiments with small systems can already lead to interesting results. The phase diagrams for other parameters with both PBC and OBC are given in Appendix~\ref{Sec:ED}, where we quantify another choice of the expectation value $\braket{\cdots}_\mathrm{LR} := \braket{\psi'_0 | \cdots | \psi_0}$ with $\bra{\psi'_0}$ being the left ground state~\cite{Yamamoto2019}. All the phase diagrams are qualitatively similar, indicating that our results do not depend strongly on the choice of the expectation values and the boundary conditions.

\section{Discussion}

Here we have shown that a quantum many-body system can undergo activity-induced phase transition in a similar manner as in the classical MIPS but with a richer phase diagram.
The fact that the addition of a simple spin-dependent hopping can lead to non-trivial phases indicates the potential of open quantum systems.
Models with asymmetric hopping have been studied extensively in the recent context of non-Hermitian topological phases~\cite{Murugan2017, Gong2018}.
It will be interesting to consider the topological characterization of phases in strongly interacting systems such as in the model studied here.
Furthermore, the correspondence between the quantum Hamiltonian and the classical transition rate matrix with bias indicates that dynamical phase transitions in general classical kinetics can in principle be probed by zero-temperature phase transitions in quantum experiments.
This connection is so far restricted to a stoquastic Hamiltonian (i.e., matrix with all its off-diagonal terms being real non-positive); exploring other models of quantum active matter, especially non-stoquastic models that have no classical analogs, will be an interesting next step.

\acknowledgements

We thank Shin-ichi Sasa, Masato Itami, Hiroyoshi Nakano, Tomohiro Soejima, Masaya Nakagawa, Yuto Ashida, and Hosho Katsura for the scientific discussions. We are also thankful to Zongping Gong, Takahiro Nemoto, Takaki Yamamoto, and Yoshihiro Michishita for helpful comments. The numerical calculations have been performed on cluster computers at RIKEN iTHEMS. 
K.A. is supported by JSPS KAKENHI Grant No. JP20K14435, and the Interdisciplinary Theoretical and Mathematical Sciences Program (iTHEMS) at RIKEN. K.T. is supported by the U.S. Department of Energy (DOE), Office of Science, Basic Energy Sciences (BES), under Contract No. AC02-05CH11231 within the Ultrafast Materials Science Program (KC2203). K.T. also thanks JSPS for support from Overseas Research Fellowship. K.K is supported by JSPS KAKENHI Grants No. JP18H04760, No. JP18K13515, No. JP19H05275, and No. JP19H05795.

\appendix

\section{Mapping to the classical model}
\label{App:Mapping}

We will show that the Hamiltonian \eqref{Eq:Hamiltonian} is mapped to the active lattice gas model (\CModel{}) under the classical condition ($U_1 = 2 J$ and $U_2 = \varepsilon J$).
First, defining $W := -H (U_1 = 2J, U_2 = \varepsilon J)$, we can obtain
\begin{align}
W =& \hat{P} \bigg\{ J \sum_{\braket{i, j}, s} ( a_{i, s}^\dag a_{j, s} + a_{j, s}^\dag a_{i, s} ) \nonumber \\
& + \varepsilon J \sum_{i, s} s ( a_{i, s}^\dag a_{i - \hat{x}, s} - a_{i, s}^\dag a_{i + \hat{x}, s} ) \nonumber \\
& + h \sum_{i, s} a_{i, s}^\dag a_{i, -s} - J \sum_{\braket{i, j}, s} [ \hat{n}_{i, s} (1 - \hat{n}_{j} ) + \hat{n}_{j, s} (1 - \hat{n}_{i} ) ] \nonumber \\
& - \varepsilon J \sum_{i, s} s [ \hat{n}_{i, s} (1 - \hat{n}_{i + \hat{x}} ) - \hat{n}_{i, s} ( 1 - \hat{n}_{i - \hat{x}} ) ] - h \sum_{i, s} \hat{n}_{i, s} \bigg\} \hat{P}.
\label{Eq:HamiltonianMethods}
\end{align}
Here, we explicitly introduce the projection operator $\hat{P}$ to a partial Fock space where the total particle number is $N$ with no multiple occupancy.

Using $W_{\mathcal{C}, \mathcal{C}'} := \braket{\mathcal{C} | W | \mathcal{C}'}$, where $\ket{\mathcal{C}}$ is the Fock-space basis corresponding to a $N$-particle configuration $\mathcal{C}$ ($:= \{ n_{i, s} \}$), we can show that (i) $\sum_\mathcal{C} W_{\mathcal{C}, \mathcal{C}'} = 0$ and (ii) $W_{\mathcal{C}, \mathcal{C}'} \geq 0$ for $\mathcal{C} \neq \mathcal{C}'$.
Thus, we can think of $W_{\mathcal{C}, \mathcal{C}'}$ as a transition rate matrix of a classical Markov process 
which yields the master equation:
\begin{equation}
\frac{d P (\mathcal{C}, t)}{d t} = \sum_{\mathcal{C}'} W_{\mathcal{C}, \mathcal{C}'} P (\mathcal{C}', t).
\label{Eq:MasterEq}
\end{equation}
where $P(\mathcal{C}, t)$ is the probability of configuration $\mathcal{C}$ at time $t$.
The first three terms of \eqref{Eq:HamiltonianMethods} (non-diagonal elements of $W_{\mathcal{C}, \mathcal{C}'}$) represent a symmetric hopping rate, a spin-dependent change in the hopping rate, and a spin flipping rate; the last three terms of \eqref{Eq:HamiltonianMethods} (diagonal elements of $W_{\mathcal{C}, \mathcal{C}'}$) represent the corresponding escape rates.

Using a state vector $\ket{\psi (t)} = \sum_\mathcal{C} P (\mathcal{C}, t) \ket{\mathcal{C}}$ according to the Doi-Peliti method~\cite{Doi1976a,Peliti1985}, we can find that \eqref{Eq:MasterEq} is nothing but the imaginary-time Schr{\" o}dinger equation, $d \ket{\psi (t)} / d t = - H (U_1 = 2 J, U_2 = \varepsilon J) \ket{\psi (t)}$.
Thus, the steady-state of the \CModel{} represented by $\ket{\psi (t \to \infty)}$ is equivalent to the ground state of the Hamiltonian, $\ket{\psi_0}$.
Also, using the coherent state $\bra{P} = \bra{0} \exp (\sum_{i, s} a_{i, s})$, we can express the expectation value of a classical physical quantity $A (\{ n_{i, s} \})$ as $\braket{A} (t) = \sum_\mathcal{C} A (\mathcal{C}) P (\mathcal{C}, t) = \braket{P | A (\{ \hat{n}_{i, s} \}) | \psi (t)}$.
Especially for the steady-state ($t \to \infty$), we obtain $\braket{A} (t \to \infty) = \braket{P | A (\{ \hat{n}_{i, s} \}) | \psi_0} = \braket{A}_\mathrm{C}$.

\section{Details of analytical and numerical results for ALG}

\subsection{Monte Carlo simulation}
\label{App:MC}

Setting a time step $\Delta t$ [$= O(N^{-1})$], we first randomly choose a particle from $N$-particles.
Then, we flip the particle's spin from $s$ to $-s$ with probability $h N \Delta t$ or move the particle to a neighboring empty site with probability $s (1 + \varepsilon) J N \Delta t$, $s (1 - \varepsilon) J N \Delta t$, or $J N \Delta t$ depending on the hopping direction.
We repeat this procedure $M$ $(:= m N)$ times, which we call $m$-MC steps, until the total time $T$ ($:= m N \Delta t$) is reached.

\begin{figure*}[t]
\centering
\includegraphics[scale=0.8]{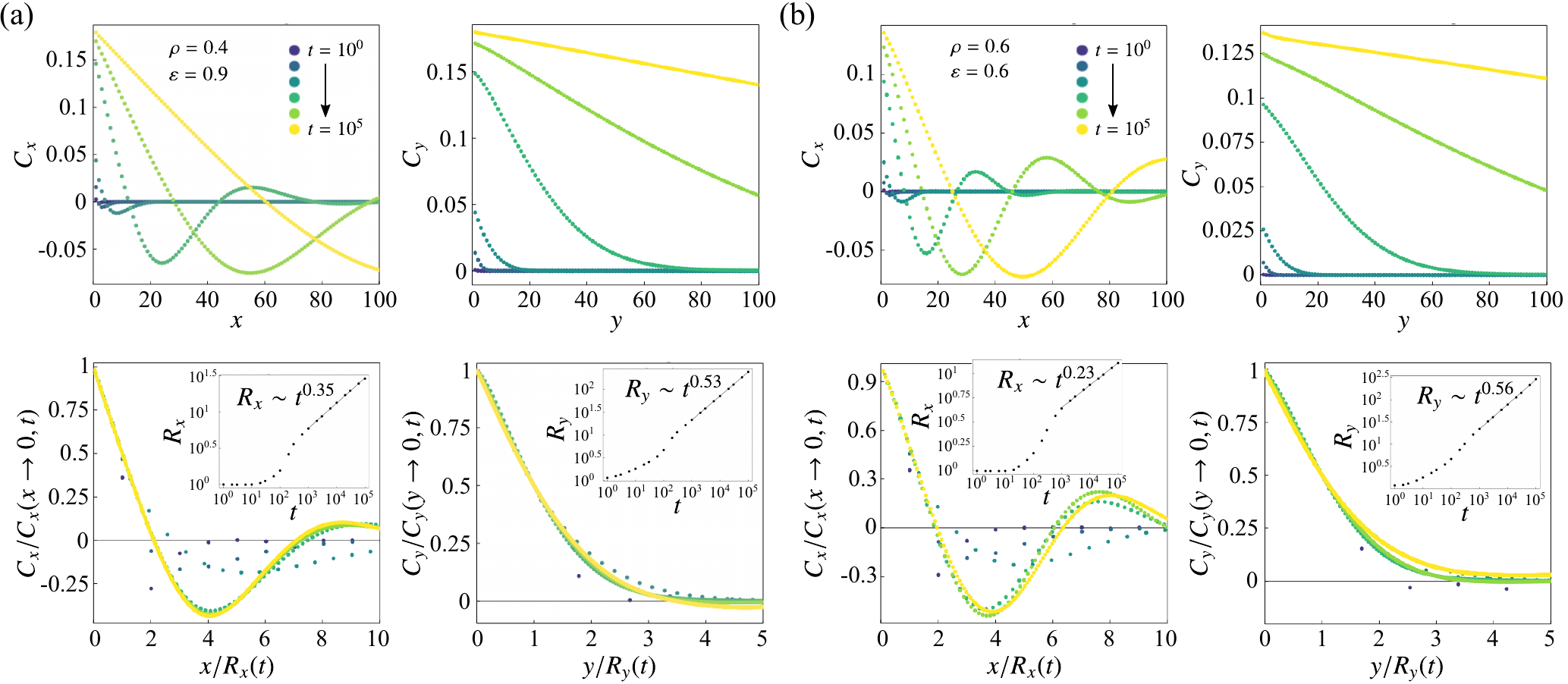}
\caption{(a) Space/time-dependence of the density correlation functions $C_x$ and $C_y$ (upper figures) and their rescaled plots (lower figures) with the time-dependence of the typical domain sizes $R_x$ and $R_y$, for $\rho = 0.4$ and $\varepsilon = 0.9$.
(b) Similar plots to (a) for $\rho = 0.6$ and $\varepsilon = 0.6$.
For all figures, we set $L_x = L_y = 3200$.
See Figs.~\ref{Fig:CMIPS_PS}(b) and (c) for $\rho = 0.6$ and $\varepsilon = 0.9$.}
\label{Fig:CMIPS_PS_app}
\end{figure*}

In simulations, we took $\Delta t = 1 / [N (4 J + h)]$ with $h = 0.025J$.
For Figs.~\ref{Fig:CMIPS}(b) and (c), we used $m = 2 \times 10^6$, ran 50 independent simulations, and took $51$ samples from each simulation for averaging.
For Figs.~\ref{Fig:CMIPS_homo}(a) and (c), we used $m = 10^6$ and ran 12000 independent simulations for averaging.
For Figs.~\ref{Fig:CMIPS_PS}(b) and (c) as well as Fig.~\ref{Fig:CMIPS_PS_app}, we ran 10 independent simulations for averaging.
We explain the details of simulations for Fig.~\ref{Fig:CMIPS_FSS} in Appendix~\ref{App:FSS}.
In all simulations, we set the disordered state with no spatial correlation as the initial state.

\begin{figure*}[t]
\centering
\includegraphics[scale=0.8]{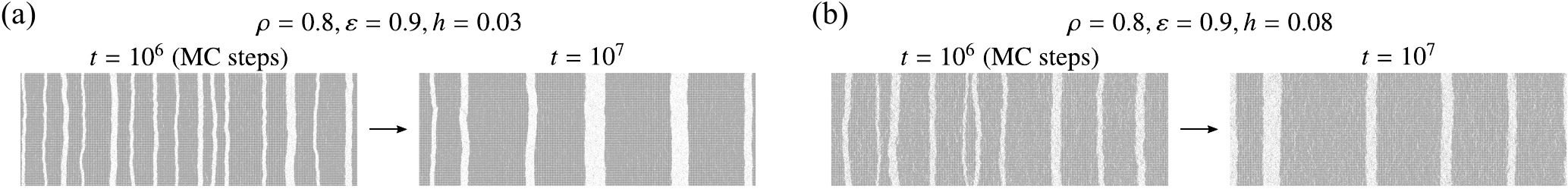}
\caption{Typical time evolution in systems with $(L_x, L_y) = (1200, 400)$ for (a) $\rho = 0.8$, $\varepsilon = 0.9$, and $h = 0.03$; (b) $\rho = 0.8$, $\varepsilon = 0.9$, and $h = 0.08$.}
\label{Fig:CMIPS_PS_app2}
\end{figure*}

\subsection{Langevin equation for spin-density field}
\label{App:Pathintegral}

Considering the ALG, we can obtain the probability density for a dynamical path of configurations $\{ n_{j, s} ( t ) \}_{t \in [ 0, T ]}$, where $n_{j,s} (t)$ is the occupancy of the site $j$ and spin $s$ at time $t$, as~\cite{Lefevre2007}
\begin{equation}
P [ n_{j, s} ] = \int D \tilde{n}_{j, s} \exp ( {- S [ n_{j, s}, \tilde{n}_{j, s} ]} ),
\label{Eq:ProbALG}
\end{equation}
where $\int D \tilde{n}_{j, s} ( \cdots )$ is the functional integral over all the possible dynamical paths of the conjugate field $\{ \tilde{n}_{j, s} ( t ) \}_{t \in [ 0, T ]}$.
Here, the action $S$ is given as
\begin{align}
S := & - i \int_0^T d t \sum_{j, s} \tilde{n}_{j, s} \partial_t n_{j, s} \nonumber \\
& - \int_0^T d t \sum_{j, s} n_{j, s} \left\{ J \sideset{}{^{( j )}} \sum_k ( 1 - n_k ) \left[ e^{i ( \tilde{n}_{j, s} - \tilde{n}_{k, s} )} - 1 \right] \right. \nonumber \\
& \left. + h \left[ e^{i ( \tilde{n}_{j, s} - \tilde{n}_{j, - s} )} - 1 \right] \right\} \nonumber \\
& - \int_0^T d t \sum_{j, s} n_{j, s} J \varepsilon s \left\{ ( 1 - n_{j + \hat{x}} ) \left[ e^{i ( \tilde{n}_{j, s} - \tilde{n}_{j + \hat{x}, s} )} - 1 \right] \right. \nonumber \\
& \left. - ( 1 - n_{j - \hat{x}} ) \left[ e^{i ( \tilde{n}_{j, s} - \tilde{n}_{j - \hat{x}, s} )} - 1 \right] \right\},
\label{Eq:ActionALG}
\end{align}
where $\sum_k^{(j)} (\cdots)$ is the summation over the sites adjacent to the site $j$.
See Ref.~\cite{Partridge2019} for the similar path-integral formulation~\cite{Lefevre2007} applied to the isotropic ALG.

Assuming that $\rho_s (\bm{r}, t)$ [$:= n_{j, s} (t)$] and $\tilde{\rho}_s (\bm{r}, t)$ [$:= \tilde{n}_{j, s} (t)$] are slowly varying on a scale of the lattice constant $a$, we approximate the action $S$ up to $O ( a^2 )$.
We also discard $O ( ( \tilde{n}_{j, s} - \tilde{n}_{j, -s} )^3 )$ and higher-order terms to consider only the Gaussian noise in the resulting Langevin equation.
Then, we can rewrite the action as $S \simeq S^{(1)}_\mathrm{cont}[\rho_s, \tilde{\rho}_s] + S^{(2)}_\mathrm{cont}[\rho_s, \tilde{\rho}_s]$, where
\begin{widetext}
\begin{equation}
S^{(1)}_\mathrm{cont} := - i \int_0^T d t \int \frac{d^2 \bm{r}}{a^2} \sum_{s} \tilde{\rho}_s \left\{ \partial_t \rho_s - J a^2 \left( \nabla^2 \rho_s - \rho_{- s} \nabla^2 \rho_s + \rho_s \nabla^2 \rho_{-s} \right) + 2 a \varepsilon J s \partial_x \left[ \left( 1 - \rho_+ - \rho_- \right) \rho_s \right] + h s (\rho_+ - \rho_-) \right\},
\end{equation}
and
\begin{equation}
S^{(2)}_\mathrm{cont} := \frac{1}{2} \int_0^T d t \int \frac{d^2 \bm{r}}{a^2} \left[ \sum_s 2 J a^2 \left( 1 - \rho_+ - \rho_- \right) \rho_s \left( \nabla \tilde{\rho}_s \right)^2 + h (\rho_+ + \rho_-) \left( \tilde{\rho}_+ - \tilde{\rho}_- \right)^2 \right].
\label{Eq:ActionALGCont}
\end{equation}
\end{widetext}
Introducing noise variables $\xi_s (\bm{r}, t)$, we transform the path probability $P [\rho_s]$, which is the continuum counterpart of $P[n_{i,s}]$, as~\cite{Lefevre2007}
\begin{equation}
P [ \rho_s ] = \int D \tilde{\rho}_s D \xi_s \exp \left( {- S_\mathrm{cont}^{(1)} [ \rho_s, \tilde{\rho}_s ] -S_\mathrm{cont}^{(2)'} [ \rho_s, \tilde{\rho}_s, \xi_s ] } \right),
\label{Eq:ProbALGCont}
\end{equation}
where
\begin{equation}
S_\mathrm{cont}^{(2)'} := \int_0^T d t \int \frac{d^2 \bm{r}}{a^2} \left[ \frac{1}{2} \sum_{s, s'} \xi_s ( M^{-1} )_{s, s'} \xi_{s'} + i \sum_s \tilde{\rho}_s \xi_s \right].
\end{equation}
Here, $M_{s, s'}$ is a differential operator given by $M_{s, s'} := \delta_{s, s'} [- 2 J \nabla \cdot (1 - \rho_+ - \rho_-) \rho_s \nabla] + (2 \delta_{s, s'} - 1) h (\rho_+ + \rho_-)$.
Following the approach developed by Martin, Siggia, Rose, Janssen, and de Dominicis (MSRJD)~\cite{Martin1973, Janssen1976, Dominicis1976}, we can obtain the Langevin equation that is equivalent to Eq.~\eqref{Eq:ProbALGCont} as
\begin{align}
\partial_t \rho_s = & J (\nabla^2 \rho_s - \rho_{-s} \nabla^2 \rho_s + \rho_s \nabla^2 \rho_{-s}) \nonumber \\
& - 2 s \varepsilon J \partial_x [(1 - \rho_+ - \rho_-) \rho_s] - h (\rho_s - \rho_{-s}) + \xi_s,
\label{Eq:CGLangevin_App}
\end{align}
where we set $a = 1$, $\xi_s (\bm{r}, t)$ is the Gaussian white noise with $\braket{\xi_s (\bm{r}, t)} = 0$, and $\braket{\xi_s (\bm{r}, t) \xi_{s'} (\bm{r}', t')} = a^2 \delta (t - t') M_{s, s'} \delta (\bm{r} - \bm{r}')$.
This equation describes the stochastic dynamics of the coarse-grained variables $\rho_s (\bm{r}, t)$.

\begin{figure*}[t]
\centering
\includegraphics[scale=0.8]{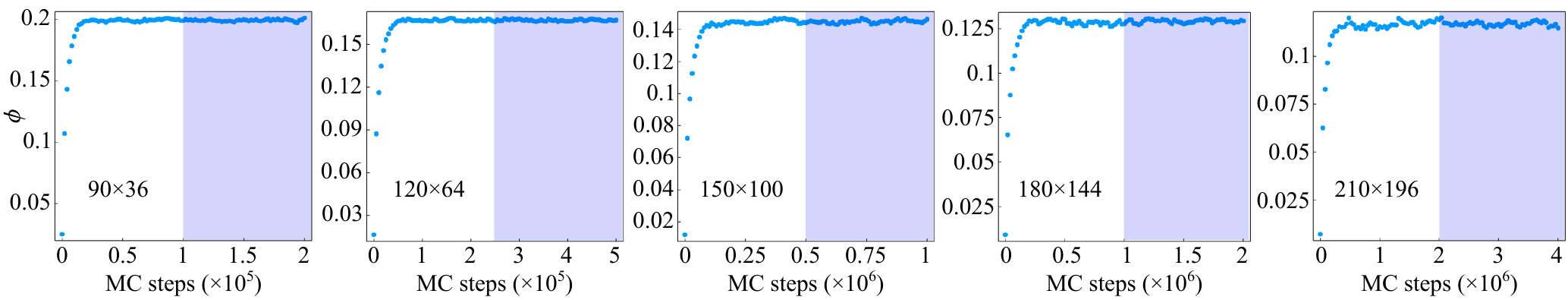}
\caption{Time-dependence of the order parameter $\phi$ averaged over independent simulations for $\rho = 0.6$ and $\varepsilon = 0.362 \ (\simeq \varepsilon_{c})$.
The number of performed independent simulations is 10000 for $(L_x, L_y) = (90, 36)$ and $(120, 64)$; 4000 for $(L_x, L_y) = (150, 100)$; 2000 for $(L_x, L_y) = (180, 144)$; 1000 for $(L_x, L_y) = (210, 196)$.
For further averaging, we took 51 samples from the time region with purple color at equal intervals.}
\label{Fig:CMIPS_FSS_app1}
\end{figure*}

\begin{figure}[t]
\centering
\includegraphics[scale=0.8]{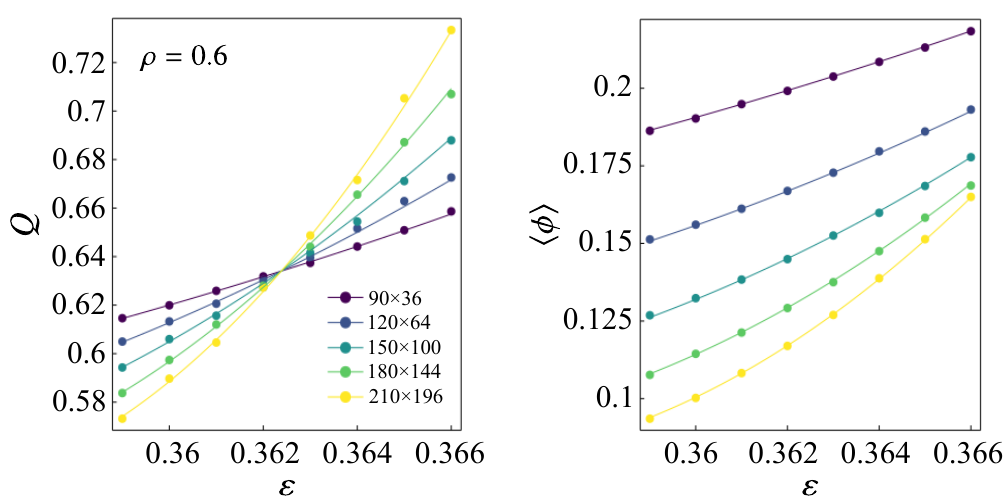}
\caption{The Binder cumulant $Q$ and the order parameter $\braket{\phi}$ obtained from simulations (colored dots), which correspond to Figs.~\ref{Fig:CMIPS_FSS}(a) and (b), and the best-fitted curves (colored lines).}
\label{Fig:CMIPS_FSS_app2}
\end{figure}

\subsection{Linearization of Langevin equation}
\label{App:Linearization}

Defining the total density $\rho_\mathrm{tot} (\bm{r}, t) := \rho_+ (\bm{r}, t) + \rho_- (\bm{r}, t)$ and the magnetization $m (\bm{r}, t) := \rho_+ (\bm{r}, t) - \rho_- (\bm{r}, t)$, we can rewrite Eq.~\eqref{Eq:CGLangevin} [or \eqref{Eq:CGLangevin_App}] as
\begin{equation}
\partial_t \rho_\mathrm{tot} = J \nabla^2 \rho_\mathrm{tot} - 2 \varepsilon J \partial_x [(1 - \rho_\mathrm{tot}) m] + \xi_\rho
\label{Eq:Langevin_sigma}
\end{equation}
and
\begin{equation}
\partial_t m = J [(1 - \rho_\mathrm{tot}) \nabla^2 m + m \nabla^2 \rho_\mathrm{tot}] - 2 \varepsilon J \partial_x [(1 - \rho_\mathrm{tot}) \rho_\mathrm{tot}] - 2 h m + \xi_m.
\label{Eq:Langevin_m}
\end{equation}
Here, $\xi_\rho (\bm{r}, t) := \xi_+ (\bm{r}, t) + \xi_- (\bm{r}, t)$ and $\xi_m (\bm{r}, t) := \xi_+ (\bm{r}, t) - \xi_- (\bm{r}, t)$.
Since $m$ is a fast mode which decays exponentially according to the $-2 h m$ term in Eq.~\eqref{Eq:Langevin_m}, we can set $\partial_t m = 0$ to examine long-time evolution of $\rho_\mathrm{tot}$, which is a slow mode due to the particle number conservation.

Using the density fluctuation $\varphi (\bm{r}, t) := \rho_\mathrm{tot} (\bm{r}, t) - \rho$, we can rewrite Eq.~\eqref{Eq:Langevin_sigma} as
\begin{equation}
\partial_t \varphi = J \nabla^2 \varphi - 2 \varepsilon J (1 - \rho) \partial_x m + 2 \varepsilon J \partial_x (\varphi m) + \xi_\rho.
\label{Eq:Langevin_phi}
\end{equation}
Setting $\partial_t m = 0$ in Eq.~\eqref{Eq:Langevin_m}, we can linearize $m (\bm{r}, t)$ with respect to $\varphi (\bm{r}, t)$ as
\begin{equation}
m \simeq [2 h - J (1 - \rho) \nabla^2]^{-1} [2 \varepsilon J (2 \rho - 1) \partial_x \varphi + \xi_m],
\label{Eq:Linmeq}
\end{equation}
where we neglect the $\varphi$-dependence of the noise~\cite{Demery2014,Poncet2017,Poncet2021}.
Substituting Eq.~\eqref{Eq:Linmeq} into Eq.~\eqref{Eq:Langevin_phi}, we can obtain the linearized equation of $\varphi (\bm{r}, t)$ as
\begin{align}
\partial_t \varphi \simeq & J \nabla^2 \varphi - 4 \varepsilon^2 J^2 (1 - \rho) (2 \rho - 1) \nonumber \\
& \times [2 h - J (1 - \rho) \nabla^2]^{-1} {\partial_x}^2 \varphi + \xi_\varphi,
\label{Eq:Linphieq}
\end{align}
where $\xi_\varphi := \xi_\rho - 2 \varepsilon J (1 - \rho) [2 h - J (1 - \rho) \nabla^2]^{-1} \partial_x \xi_m$, $\braket{\xi_\varphi (\bm{r}, t)} = 0$, and
\begin{align}
& \braket{\xi_\varphi (\bm{r}, t) \xi_\varphi (\bm{r}', t')} = - 2 J (1 - \rho) \rho \nabla^2 \delta (\bm{r} - \bm{r}') \delta (t - t') \nonumber \\
& - 8 \varepsilon^2 J ^2 (1 - \rho)^2 \rho [2 h - J (1 - \rho) \nabla^2]^{-1} {\partial_x}^2 \delta (\bm{r} - \bm{r}') \delta (t - t').
\label{Eq:noisephi}
\end{align}

Applying Fourier transformation, $\varphi (\bm{k}, t) := \int d^2 \bm{r} \exp (- i \bm{k} \cdot \bm{r}) \varphi (\bm{r}, t)$, we can solve Eq.~\eqref{Eq:Linphieq} and finally obtain the structure factor, $S_\mathrm{lin} (\bm{k}) := (L_x L_y)^{-1} \lim_{t \to \infty} \braket{|\varphi (\bm{k}, t)|^2}$, as
\begin{align}
& S_\mathrm{lin} (\bm{k}) = (1 - \rho) \rho \nonumber \\
& \times \frac{ [2 h + J (1 - \rho) \bm{k}^2] \bm{k}^2 + 4 \varepsilon^2 J (1 - \rho) {k_x}^2}{[2 h + J (1 - \rho) \bm{k}^2] \bm{k}^2 - 4 \varepsilon^2 J (1 - \rho) (2 \rho - 1) {k_x}^2}.
\end{align}

\subsection{Anisotropic growth in PS state}
\label{App:Growth}

In Fig.~\ref{Fig:CMIPS_PS_app}, We show the space and time-dependence of the density correlation function for $(\rho, \varepsilon) = (0.4, 0.9)$ and $(0.6, 0.6)$, which are different from the parameters used for Fig.~\ref{Fig:CMIPS_PS}.
We can see that the anisotropic power law of the typical domain size, $R_x (t) \sim t^{\alpha_x}$ and $R_y (t) \sim t^{\alpha_y}$ with $\alpha_x < \alpha_y$, holds both for $(\rho, \varepsilon) = (0.4, 0.9)$ and $(0.6, 0.6)$ as observed for $(\rho, \varepsilon) = (0.6, 0.9)$ (Fig.~\ref{Fig:CMIPS_PS}), though the exponent seems non-universal.

To examine whether the counterpart of the bubbly phase separation, which has been observed in the isotropic ALG~\cite{Shi2020}, can appear in the anisotropic ALG, we performed simulations using systems with $(L_x, L_y) = (1200, 400)$.
We did not find evidence of an analog of the bubbly phase separation, where the bubbles of the low-density phase should be nucleated inside the bulk high-density phase, though the steady-state has not been reached by the end of the simulation ($10^7$ MC steps) [see Figs.~\ref{Fig:CMIPS_PS_app2}(a) and (b) for typical snapshots for $(\rho, \varepsilon, h) = (0.8, 0.9, 0.03)$ and $(0.8, 0.9, 0.08)$, respectively].

\begin{figure*}[t]
\centering
\includegraphics[scale=0.8]{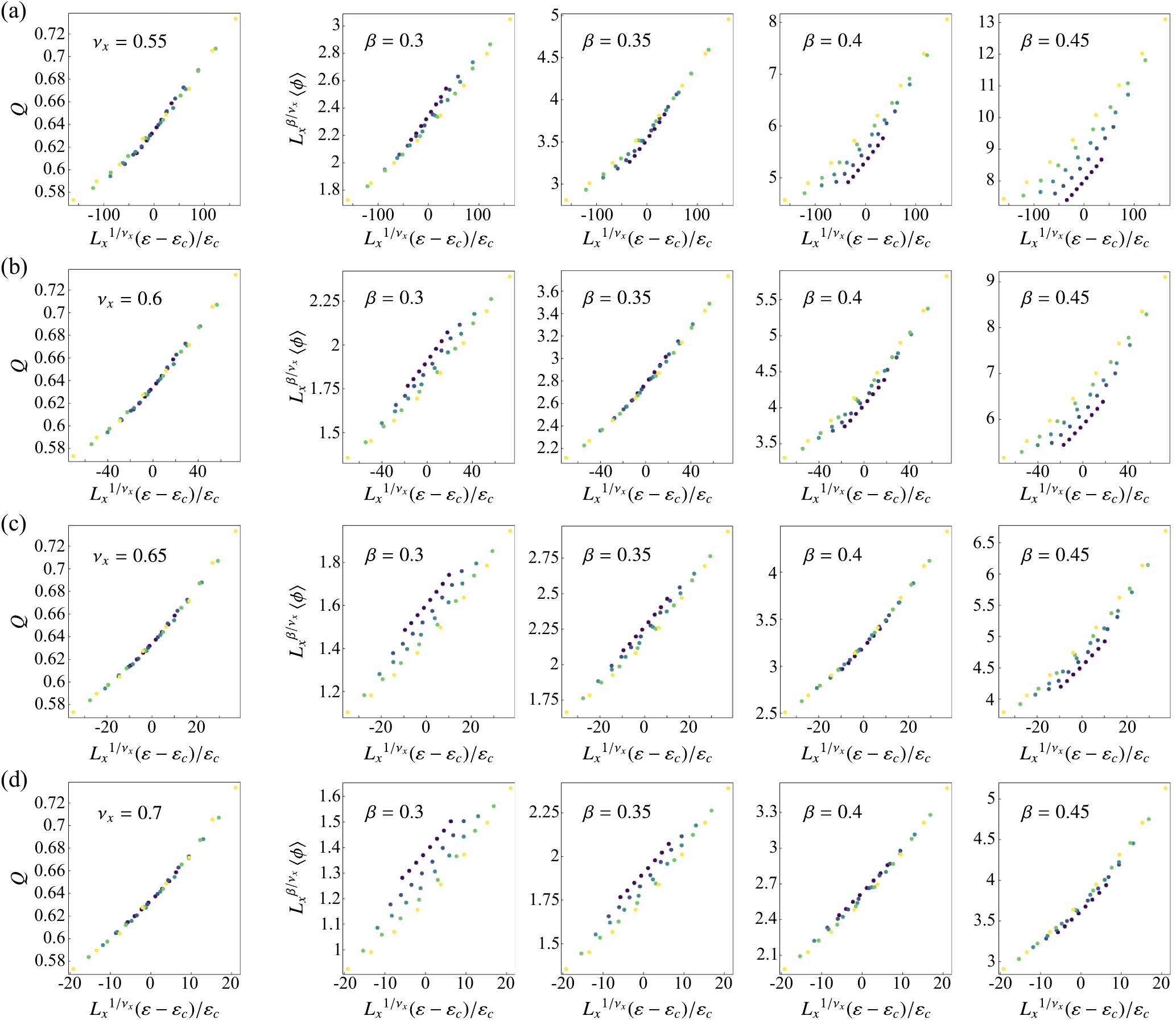}
\caption{Rescaled $Q$ and $\braket{\phi}$ for the best-fitted parameters with $\nu_x$ and $\beta$ fixed: (a) $\nu_x = 0.55$ and $\beta = 0.3$--0.45, (b) $\nu_x = 0.6$ and $\beta = 0.3$--0.45, (c) $\nu_x = 0.65$ and $\beta = 0.3$--0.45, and (d) $\nu_x = 0.7$ and $\beta = 0.3$--0.45.
We used the same simulation data as in Fig.~\ref{Fig:CMIPS_FSS}, and the best-fitted critical point is $\varepsilon_{c} \simeq 0.362$ for (a) through (d).}
\label{Fig:CMIPS_FSS_app3}
\end{figure*}

\begin{figure}[t]
\centering
\includegraphics[scale=0.8]{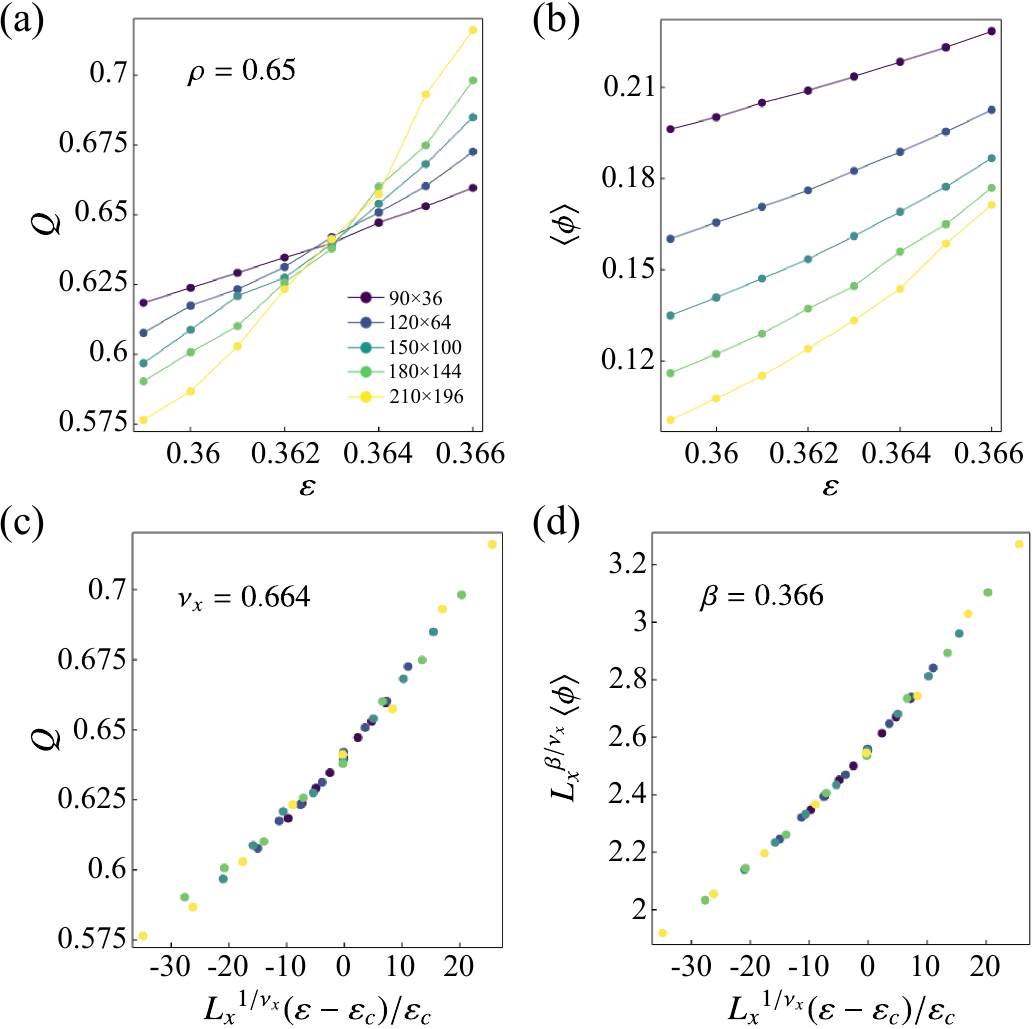}
\caption{Counterparts of Fig.~\ref{Fig:CMIPS_FSS} for $\rho = 0.65$ with $L_y / {L_x}^2 = 1 / 15^2$.
(a) $\varepsilon$ and system size-dependence of the Binder ratio $Q$.
The solid lines are guides for the eyes.
(b) $\varepsilon$ and system size-dependence of $\braket{\phi}$.
(c) $Q$ as a function of the rescaled $\varepsilon$ with the best-fitted $\varepsilon_{c}$ $(\simeq 0.363)$ and $\nu_x$ $(\simeq 0.664)$.
(d) Rescaled $\braket{\phi}$ as a function of the rescaled $\varepsilon$ with the best-fitted $\beta$ $(\simeq 0.366)$ and the same values of $\varepsilon_{c}$ and $\nu_x$ as (c).}
\label{Fig:CMIPS_FSS_app4}
\end{figure}

\subsection{Effective model for critical dynamics}
\label{App:Criticalmodel}

Setting $\partial_t m = 0$ by focusing on long-time evolution and using $\varphi (\bm{r}, t) = \rho_\mathrm{tot} (\bm{r}, t) - \rho$ as in Appendix~\ref{App:Linearization}, we can iteratively solve Eq.~\eqref{Eq:Langevin_m} as
\begin{widetext}
\begin{align}
m &= \frac{1}{2h} [2 \varepsilon J (2 \rho - 1) \partial_x \varphi + 2 \varepsilon J \partial_x \varphi^2 + \xi_m] + \frac{J}{2h} \left[ (1 - \rho) \nabla^2 m + \nabla \cdot (-\varphi \nabla m + m \nabla \varphi) \right] \nonumber \\
&= \frac{1}{2h} [2 \varepsilon J (2 \rho - 1) \partial_x \varphi + 2 \varepsilon J \partial_x \varphi^2 + \xi_m] + \frac{J}{4h^2} (1 - \rho) \left[ 2 \varepsilon J (2 \rho - 1) \nabla^2 \partial_x \varphi \right] + O(\nabla^4 \partial_x \varphi, \nabla^2 \partial_x \varphi^2, \nabla^2 \xi_m).
\label{Eq:meqeff}
\end{align}
Substituting Eq.~\eqref{Eq:meqeff} into Eq.~\eqref{Eq:Langevin_phi}, we can obtain
\begin{align}
\partial_t \varphi = & J \left[ 1 - \frac{2 \varepsilon^2 J}{h} (1 - \rho) (2 \rho - 1) \right] {\partial_x}^2 \varphi + J {\partial_y}^2 \varphi - \frac{\varepsilon^2 J^3}{h^2} (1 - \rho)^2 (2 \rho - 1) {\partial_x}^4 \varphi + \frac{\varepsilon^2 J^2}{h} (4 \rho - 3) {\partial_x}^2 \varphi^2 + \frac{4 \varepsilon^2 J^2}{3 h} {\partial_x}^2 \varphi^3 \nonumber \\
& + \sqrt{2 J (1 - \rho) \rho \left[ 1 + \frac{2 \varepsilon^2 J}{h} (1 - \rho) \right]} \partial_x \eta + O({\partial_x}^2 {\partial_y}^2 \varphi, {\partial_x}^6 \varphi, {\partial_x}^4 \varphi^2, \partial_y \eta, {\partial_x}^2 \eta, \sqrt{\varphi} \partial_x \eta),
\label{Eq:Langevin_phi_nl}
\end{align}
\end{widetext}
where $\eta (\bm{r}, t)$ satisfies $\braket{\eta (\bm{r}, t)} = 0$ and $\braket{\eta (\bm{r}, t) \eta (\bm{r}', t')} = \delta (\bm{r} - \bm{r}') \delta (t -t')$.
Note that, neglecting the noise $\eta$, we can obtain the spinodal line $\varepsilon_\mathrm{sp} (\rho)$ from $1 = 2 \varepsilon_\mathrm{sp} (\rho)^2 J (1 - \rho) (2 \rho - 1) / h$ and the mean-field critical point as $\rho_{c}^\mathrm{MF} = 3 / 4$ and $\varepsilon_{c}^\mathrm{MF} = \pm 2 \sqrt{h/J}$.

Applying the MSRJD approach to Eq.~\eqref{Eq:Langevin_phi_nl}, we can show that the probability density for a dynamical path of configurations $\{ \varphi (t) \}_{t \in [0, T]}$ is given by
\begin{equation}
P[\varphi] = \int D (i \tilde{\varphi}) \exp ( -S_\varphi [\varphi, \tilde{\varphi}] ).
\end{equation}
Here, the action is given by
\begin{align}
S_\varphi := & \int_0^T d t \int d^2 \bm{r} [\tilde{\varphi} ( \partial_t \varphi - \tau_x {\partial_x}^2 \varphi - \tau_y {\partial_y}^2 \varphi + a {\partial_x}^4 \varphi \nonumber \\
& - \varv {\partial_x}^2 \varphi^2 - u {\partial_x}^2 \varphi^3 ) + c \tilde{\varphi} {\partial_x}^2 \tilde{\varphi} + (\mathrm{h.o.t.})],
\label{Eq:Action_phi}
\end{align}
where we generalize the coupling constants for each term in Eq.~\eqref{Eq:Langevin_phi_nl} as $J [1 - 2 \varepsilon^2 J (1 - \rho) (2 \rho - 1) / h] \to \tau_x$, $J \to \tau_y$, $\varepsilon^2 J^3 (1 - \rho)^2 (2 \rho - 1) / h^2 \to a$, $\varepsilon^2 J^2 (4 \rho - 3) / h \to \varv$, $4 \varepsilon^2 J^2 / (3 h) \to u$, and $J (1 - \rho) \rho [1 + 2 \varepsilon^2 J (1 - \rho) / h] \to c$.
In Eq.~\eqref{Eq:Action_phi}, $\mathrm{(h.o.t.)}$ corresponds to the higher-order terms in Eq.~\eqref{Eq:Langevin_phi_nl}, which are irrelevant in the renormalization group (RG) sense, as shown below.

We consider the tree-level RG analysis for Eq.~\eqref{Eq:Action_phi}.
Considering the scale transformation $x \to b^{-1} x$ ($b > 1$) and requiring the invariance of $\tau_y$, $a$, and $c$ under the transformation, we can obtain the scaling of other quantities as
\begin{equation}
\left\{
\begin{array}{l}
y \to b^{-2} y \\
t \to b^{-4} t \\
\varphi \to b^{1/2} \varphi \\
\tilde{\varphi} \to b^{5/2} \tilde{\varphi} \\
\tau_x \to b^2 \tau_x \\
\varv \to b^{3/2} \varv \\
u \to b u,
\end{array}
\right.
\label{Eq:RGscaling}
\end{equation}
suggesting that $\tau_x$, $\varv$, and $u$ are relevant variables.
In particular, $\tau_x \propto (\varepsilon - \varepsilon_{c})$ around the critical point.
Further, we can write each term of $(\mathrm{h.o.t.})$ in Eq.~\eqref{Eq:Action_phi} as $d_{\gamma_x \gamma_y \gamma_\varphi} \tilde{\varphi} {\partial_x}^{\gamma_x} {\partial_y}^{\gamma_y} \varphi^{\gamma_\varphi}$ or $e_{\delta_x \delta_y \delta_\varphi} \tilde{\varphi} {\partial_x}^{\delta_x} {\partial_y}^{\delta_y} \varphi^{\delta_\varphi} \tilde{\varphi}$, and the scaling of the coupling constants is obtained as
\begin{equation}
\left\{
\begin{array}{l}
d_{\gamma_x \gamma_y \gamma_\varphi} \to b^{9/2 - \gamma_x - 2 \gamma_y - \gamma_\varphi / 2} d_{\gamma_x \gamma_y \gamma_\varphi} \\
e_{\delta_x \delta_y \delta_\varphi} \to b^{2 - \delta_x - 2 \delta_y - \delta_\varphi / 2} e_{\delta_x \delta_y \delta_\varphi}.
\end{array}
\right.
\label{Eq:RGscaling2}
\end{equation}
Since $\gamma_x + 2 \gamma_y + \gamma_\varphi / 2 \geq 5$ and $\delta_x + 2 \delta_y + \delta_\varphi / 2 \geq 5 / 2$, $d_{\gamma_x \gamma_y \gamma_\varphi}$ and $e_{\delta_x \delta_y \delta_\varphi}$ are irrelevant variables.

Omitting the irrelevant variables and adjusting the density $\rho$ so that $\varv = 0$ in Eq.~\eqref{Eq:Action_phi}, we obtain the effective action for the critical dynamics of the ALG, \begin{align}
S_\varphi' := & \int_0^T d t \int d^2 \bm{r} [\tilde{\varphi} ( \partial_t \varphi - \tau_x {\partial_x}^2 \varphi - \tau_y {\partial_y}^2 \varphi + a {\partial_x}^4 \varphi \nonumber \\
& - u {\partial_x}^2 \varphi^3 ) + c \tilde{\varphi} {\partial_x}^2 \tilde{\varphi}],
\label{Eq:Action_phi2}
\end{align}
which coincides with that of the randomly driven or two-temperature lattice gas model ~\cite{Schmittmann1991,Schmittmann1993,Praestgaard1994,Praestgaard2000}.

\subsection{Finite-size scaling analysis}
\label{App:FSS}

For Fig.~\ref{Fig:CMIPS_FSS}, we performed 1000--10000 independent simulations and took 51 samples from each simulation for averaging.
In Fig.~\ref{Fig:CMIPS_FSS_app1}, we show the time-dependence of the order parameter $\phi$ averaged over independent simulations for $\varepsilon = 0.362 \ (\simeq \varepsilon_{c})$ and the time region used for further averaging.
Similar time-dependence was obtained also for other values of $\varepsilon$.

To find the critical point $\varepsilon_{c}$ and the critical exponents $\nu_x$ and $\beta$ from the obtained data [Figs.~\ref{Fig:CMIPS_FSS}(a) and (b)], we performed curve fitting with a julia package LsqFit.jl.
We first fitted the data of the Binder ratio $Q(\varepsilon, L_x)$ with the formula, $Q^{(0)} + Q^{(1)} {L_x}^{1/\nu_x} (\varepsilon - \varepsilon_{c}) + Q^{(2)} {L_x}^{2/\nu_x} (\varepsilon - \varepsilon_{c})^2$, based on the second-order expansion of the scaling form $Q(\varepsilon, L_x) = F_Q ({L_x}^{1/\nu_x} (\varepsilon - \varepsilon_{c}))$.
Here, the fitting parameters are $Q^{(0)}$, $Q^{(1)}$, $Q^{(2)}$, $\varepsilon_{c}$, and $\nu_x$.
Then, using the obtained $\varepsilon_{c}$ and $\nu_x$, we fitted the data of $\braket{\phi} (\varepsilon, L_x)$ with the formula, $\phi^{(0)} {L_x}^{-\beta / \nu_x} + \phi^{(1)} {L_x}^{-\beta / \nu_x + 1 / \nu_x} (\varepsilon - \varepsilon_{c}) + \phi^{(2)} {L_x}^{-\beta / \nu_x + 2 / \nu_x} (\varepsilon - \varepsilon_{c})^2$, based on the second-order expansion of the scaling form $\braket{\phi} (\varepsilon, L_x) = {L_x}^{-\beta / \nu_x} F_1 ({L_x}^{1/\nu_x} (\varepsilon - \varepsilon_{c}))$.
Here, the fitting parameters are $\phi^{(0)}$, $\phi^{(1)}$, $\phi^{(2)}$, and $\beta$.
We show the best-fitted curves for $Q(\varepsilon, L_x)$ and $\braket{\phi} (\varepsilon, L_x)$ in Fig.~\ref{Fig:CMIPS_FSS_app2}, and the best-fitted parameters are $\varepsilon_{c} \simeq 0.36238(4)$, $\nu_x \simeq 0.65(1)$, and $\beta \simeq 0.3928(8)$ as mentioned in Sec.~\ref{Sec:CMIPS_FSS}, where the value in the bracket is the fitting error on the last significant figure.

To check the deviation of the scaling behavior against slight changes in the estimated critical exponents, we also tried another curve fitting with $\nu_x$ and $\beta$ fixed.
Here, the fitting formulas are the same as before, but the fitting parameters are ($Q^{(0)}$, $Q^{(1)}$, $Q^{(2)}$, $\varepsilon_{c}$) in fitting $Q(\varepsilon, L_x)$ and ($\phi^{(0)}$, $\phi^{(1)}$, $\phi^{(2)}$) in fitting $\braket{\phi} (\varepsilon, L_x)$.
Plotting the rescaled curves similarly to Figs.~\ref{Fig:CMIPS_FSS}(c) and (d) for several values of $\nu_x$ and $\beta$ (Fig.~\ref{Fig:CMIPS_FSS_app3}), we find that the curves seem well-scaled for a certain range of exponents, including $(\nu_x, \beta) = (0.6, 0.35)$, which, within uncertainty, coincide with those observed for the two-temperature lattice gas model~\cite{Praestgaard2000}.
Larger-scale simulations will be necessary to accurately determine the exponents numerically.

We further performed MC simulations and the finite-size scaling analysis for $\rho = 0.65$ with $S = L_y / {L_x}^2 = 1 / 15^2$ in the same way as for $\rho = 0.6$.
The counterparts of Fig.~\ref{Fig:CMIPS_FSS} for $\rho = 0.65$ are shown in Fig.~\ref{Fig:CMIPS_FSS_app4}.
The best-fitted parameters are $\varepsilon_{c} = 0.36304(7)$, $\nu_x = 0.66(2)$, and $\beta = 0.366(1)$, which are qualitatively similar to the case of $\rho = 0.6$, given the well-scaled range of exponents for $\rho = 0.6$ (Fig.~\ref{Fig:CMIPS_FSS_app3}).

\section{Diffusion Monte Carlo simulation}
\label{Sec:DMC}

For the quantum model [Eq.~\eqref{Eq:Hamiltonian}], we first divide the Hamiltonian into two parts $H = -W - D$, where $W$ is given by \eqref{Eq:HamiltonianMethods} and $D$ is diagonal in the Fock space.
To numerically calculate the quantity $\braket{A}_\mathrm{C} = \braket{P | A (\{ \hat{n}_{i, s} \}) | \psi_0} / \braket{P | \psi_0}$ for the ground state $\ket{\psi_0}$, we transform $\braket{A}_\mathrm{C}$ as
\begin{align}
\braket{A}_\mathrm{C} =& \lim_{T \to \infty} \frac{\braket{P | A (\{ \hat{n}_{i, s} \}) e^{(W + D) T} | \psi_\mathrm{ini}}}{\braket{P | e^{(W + D) T} | \psi_\mathrm{ini}}} \nonumber \\
=& \lim_{T \to \infty} \frac{\sum_{\mathcal{C}, \mathcal{C}_0} A (\mathcal{C}) \braket{\mathcal{C} | e^{(W + D) T} | \mathcal{C}_0} P_\mathrm{ini} (\mathcal{C}_0)}{\sum_{\mathcal{C}, \mathcal{C}_0} \braket{\mathcal{C} | e^{(W + D) T} | \mathcal{C}_0} P_\mathrm{ini} (\mathcal{C}_0)}
\end{align}
where $\ket{\psi_\mathrm{ini}} := \sum_\mathcal{C} P_\mathrm{ini} (\mathcal{C}) \ket{\mathcal{C}}$ with $P_\mathrm{ini} (\mathcal{C}) \geq 0$ is an arbitrary initial state.
Instead of taking $T \to \infty$, we consider a finite but large enough $T$ for the initial state to relax to the ground state.

Splitting the total time $T$ as $T = M \Delta t$ with a time step $\Delta t$ [$= O(N^{-1})$] and writing $\mathcal{C} = \mathcal{C}_M$ for convenience, we can divide the time evolution into small steps:
\begin{align}
& \braket{\mathcal{C} | e^{(W + D) T} | \mathcal{C}_0} \nonumber \\
&= \sum_{\mathcal{C}_1, \cdots, \mathcal{C}_{M - 1}} \prod_{m = 1}^{M} \braket{\mathcal{C}_{m} | e^{(W + D) \Delta t} | \mathcal{C}_{m - 1}} \nonumber \\
&\approx \sum_{\mathcal{C}_1, \cdots, \mathcal{C}_{M - 1}} \prod_{m = 1}^{M} (\delta_{\mathcal{C}_m, \mathcal{C}_{m - 1}} + W_{\mathcal{C}_m, \mathcal{C}_{m - 1}} \Delta t) (1 + D_{\mathcal{C}_{m - 1}} \Delta t ),
\label{Eq:EvoMat}
\end{align}
where $D_{\mathcal{C}} := \braket{\mathcal{C} | D | \mathcal{C}}$ and the approximation in the third line is correct up to $O (\Delta t)$.
Since $\delta_{\mathcal{C}_m, \mathcal{C}_{m - 1}} + W_{\mathcal{C}_m, \mathcal{C}_{m - 1}} \Delta t$ is a stochastic matrix for the \CModel{}, we can approximately calculate \eqref{Eq:EvoMat} by assigning the weight $\prod_{m = 1}^M (1 + D_{\mathcal{C}_{m - 1}} \Delta t )$ to the sampled path $\mathcal{C}_{0} \to \mathcal{C}_1 \to \cdots \to \mathcal{C}_M$ in the Monte Carlo (MC) simulations of the \CModel{}.

To efficiently sample the configurations that have high probability weights but rarely appear in the MC simulation, we use the re-sampling technique~\cite{Giardina2006}.
We consider a set of configurations, $\{ \mathcal{C}_m^{(i)} \}_{i = 1}^{N_\mathrm{c}}$, which evolve independently through the MC dynamics.
Correspondingly, we introduce a set of cumulative weights, $\{ w^{(i)}_m \}_{i = 1}^{N_\mathrm{c}}$, according to the paths $\{ \mathcal{C}_0^{(i)} \to \cdots \to \mathcal{C}_m^{(i)} \}_{i = 1}^{N_\mathrm{c}}$.
Whenever the effective sample size~\cite{Martino2017}, $(\sum_{i = 1}^{N_\mathrm{c}} w_m^{(i)})^2 / \sum_{i = 1}^{N_\mathrm{c}} (w_m^{(i)})^2$, becomes smaller than $0.5 N_\mathrm{c}$ during the MC dynamics, we perform re-sampling of configurations from the distribution of $\{ \mathcal{C}_m^{(i)} \}_{i = 1}^{N_\mathrm{c}}$ weighted by $\{ w^{(i)}_m \}_{i = 1}^{N_\mathrm{c}}$ and then reset the weights as $w_m^{(i)} = 1$ for all $i$.
Using the final-time configurations and weights, $\{ \mathcal{C}_M^{(i)} \}_{i = 1}^{N_\mathrm{c}}$ and $\{ w^{(i)}_M \}_{i = 1}^{N_\mathrm{c}}$, we estimate $\braket{A}_\mathrm{C}$ as
\begin{equation}
\braket{A}_\mathrm{C} \approx \frac{\sum_{i = 1}^{N_\mathrm{c}} w^{(i)}_M \, A ( \mathcal{C}^{(i)}_M )}{\sum_{i = 1}^{N_\mathrm{c}} w^{(i)}_M}.
\end{equation}

In 2D simulations, we typically took $\Delta t = 1 / [N (4 J + h)]$ and used $(N_\mathrm{c}, M) =  (5 \times 10^3, 10^5 N)$ for Fig.~\ref{Fig:QMIPS}(a); $(N_\mathrm{c}, M) =  (10^5, 10^4 N)$ for Figs.~\ref{Fig:QMIPS}(b) and \ref{Fig:ep}(a); and $(N_\mathrm{c}, M) =  (2 \times 10^4, 2 \times 10^4 N)$ for Fig.~\ref{Fig:PD}.
In 1D simulations for Fig.~\ref{Fig:ep}(b), we took $\Delta t = 1 / [N (2 J + h)]$ and $(N_\mathrm{c}, M) =  (10^5, 10^4 N)$.
In all simulations, we set the disordered state with no spatial correlation as the initial state, while we confirmed that there is no qualitative dependence on the initial state (see Appendix~\ref{Sec:convergence}).

\begin{figure*}
\centering
\includegraphics[scale=0.8]{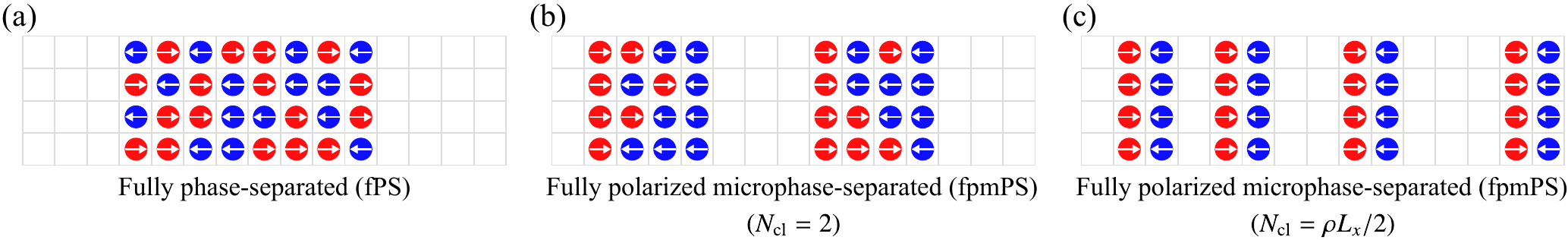}
\caption{Schematic figures of representative states.
(a) In the fPS state, a single cluster with random spins is formed and its circumference is minimized.
(b) In the fpmPS state, there are $N_\mathrm{cl}$ clusters with oppositely polarized edges.
(c) For large enough $U_2$ ($\gg J, h, U_1$), the fpmPS state is stable with the maximal number of clusters, $N_\mathrm{cl} = \rho L_x / 2$.}
\label{Fig:RefStates}
\end{figure*}

\begin{figure*}
\centering
\includegraphics[scale=0.8]{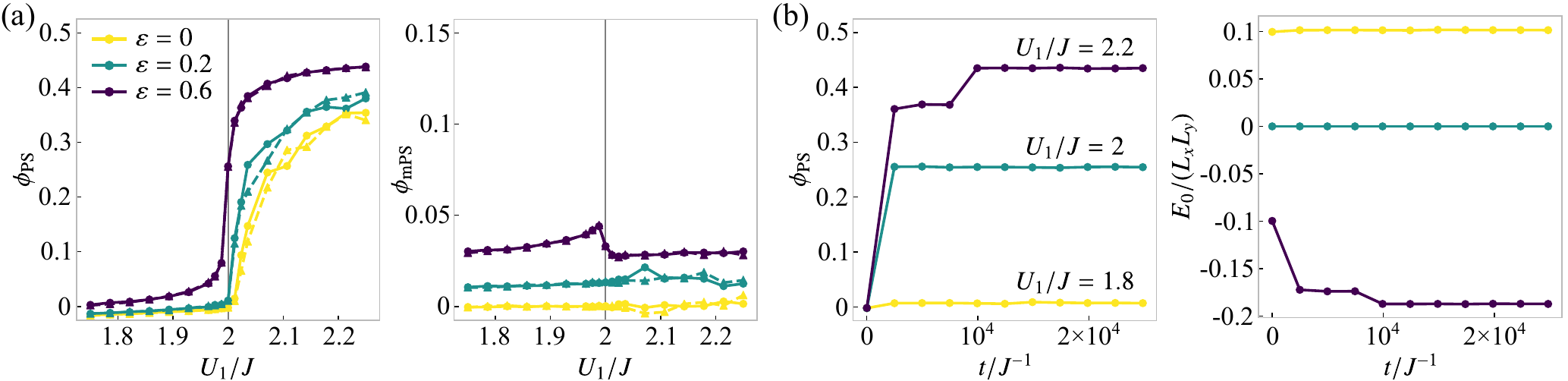}
\caption{
(a) $U_1$-dependence of the order parameters, $\phi_\mathrm{PS}$ and $\phi_\mathrm{mPS}$, obtained with the disordered (solid line with circles) or the PS (dashed line with triangles) initial state for $\varepsilon = 0, 0.2, 0.6$.
(b) Time evolution of $\phi_\mathrm{PS}$ and $E_0$ in simulations, obtained with the disordered initial state for $\varepsilon = 0.6$ and $U_1 / J = 1.8, 2, 2.2$.
In both (a) and (b), we considered $50 \times 5$ systems and used $\rho = 0.5$, $h = 0.025 J$, and $U_2 = \varepsilon J$.
Simulation parameters are $\Delta t = 1 / [N ( 4 J + h )]$, $N_\mathrm{c} = 5 \times 10^3$, and $M = 10^5 N$ as used in Fig.~\ref{Fig:QMIPS}(a) of the main text (see Appendix~\ref{Sec:DMC}).
Note that, since we set $\hbar = 1$, time and inverse of energy have the same dimension.}
\label{Fig:Conv}
\end{figure*}

\section{Properties of the model and details of the analysis}

\subsection{Generalized quantum model and classical condition}
\label{Sec:genq}

We consider a generalized version of the two-component hard-core boson model~\eqref{Eq:Hamiltonian} in the main text:
\begin{align}
& H_\mathrm{gen} = \hat{P} \left( - \sum_{i} \sum_{l = x, y} \sum_{s, r = \pm} J_{s, r}^{(l)} a_{i + r \hat{l}, s}^\dag a_{i, s}  \right. \nonumber \\
& \left. - \sum_{i} \sum_{a = 0, 1, 2, 3} \sum_{s, s' = \pm} h_a \sigma^a_{s, s'} a_{i, s}^\dag a_{i, s'} - \sum_{i} \sum_{l = x, y} \sum_{s, r = \pm} U_{s, r}^{(l)} \hat{n}_{i, s} \hat{n}_{i + r \hat{l}} \right) \hat{P},
\label{Eq:GenHamiltonian}
\end{align}
where $\sigma^0$ is the $2 \times 2$ identity matrix, $\sigma^a$ ($a = 1, 2, 3$) are the Pauli matrices, and $\hat{P}$ is the projection to a partial Fock space where the total particle number is $N$ with no multiple occupancy.
We assume $[ a_{i, s}, a_{j, s'}^\dag ] = [ a_{i, s}, a_{j, s'} ] = [ a_{i, s}^\dag, a_{j, s'}^\dag ] = 0$ for $( i, s ) \neq ( j, s' )$; $\{ a_{i, s}, a_{i, s}^\dag \} = 1$ and ${a_{i, s}}^2 = {( a_{i, s}^\dag )}^2 = 0$.
The first term of Eq.~\eqref{Eq:GenHamiltonian} represents hopping, which is, in general, non-Hermitian and dependent on the spin and/or the hopping direction.
The second and third terms represent the effect of external fields and the generalized nearest-neighbor interactions, respectively.
For $J_{s, r}^{( l )} = ( 1 + s r \varepsilon \delta_{l, x} ) J$, $h_a = - ( 4 J + h ) \delta_{a, 0} + h \delta_{a, 1}$, and $U_{s, r}^{( l )} = U_1 / 2 + s r U_2 \delta_{l, x}$, we can reproduce the model~\eqref{Eq:Hamiltonian} in the main text.

Here, we take $U_{s, r}^{( l )} = J_{s, r}^{( l )}$, $h_0 = - \sum_{l, s, r} J_{s, r}^{( l )} / 2 - h_1$, $h_2 = 0$, and $h_3 = - \sum_{l, s, r} s J_{s, r}^{( l )} / 2$ with arbitrary $J_{s, r}^{(l)} > 0$ and $h_1 > 0$, which is the generalized classical condition (see the main text and Appendix~\ref{App:Mapping}).
Defining $W := -H_\mathrm{gen}$ under this classical condition, we can obtain
\begin{align}
W = \hat{P} \left\{ \sum_{i} \sum_{l = x, y} \sum_{s, r = \pm} J_{s, r}^{(l)} \left[ a_{i + r \hat{l}, s}^\dag a_{i, s} - \hat{n}_{i, s} \left( 1 - \hat{n}_{i + r \hat{l}} \right) \right] \right. \nonumber \\
\left. + \sum_{i} \sum_{s = \pm} h_1 \left( a_{i, s}^\dag a_{i, -s} - \hat{n}_{i, s} \right) \right\} \hat{P}.
\end{align}
Defining $W_{\mathcal{C}, \mathcal{C}'} := \braket{\mathcal{C} | W  | \mathcal{C}'}$, where $\ket{\mathcal{C}}$ is the Fock-space basis, we can show that (i) $\sum_{\mathcal{C}} W_{\mathcal{C}, \mathcal{C}'} = 0$ and (ii) $W_{\mathcal{C}, \mathcal{C}'} \geq 0$ for $\mathcal{C} \neq \mathcal{C}'$, and thus we can interpret $W_{\mathcal{C}, \mathcal{C}'}$ as a transition rate matrix of a classical Markov process.
Under this interpretation, $J_{s, r}^{( l )}$ is the hopping rate of a particle with spin $s$ from a site $i$ to the adjacent site $i + r \hat{l}$, and $h_1$ is the spin flipping rate.

Lastly, we briefly discuss the quantum model~\eqref{Eq:Hamiltonian} in the main text for the open boundary condition (OBC).
OBC in a quantum system is when the hopping to the outside of the $L_x \times L_y$ region ($\Omega$) is prohibited and there are no interactions between the particles inside and the outside of $\Omega$.
This is different to the OBC in the classical system such as in ALG, meaning that there is no classical line in the case of OBC.
We conducted exact diagonalization calculations for a small 1D quantum system to check the effect of this open boundary condition on the phase diagram (see Appendix~\ref{Sec:ED} and Fig.~\ref{Fig:ED} for more details).
On the other hand, we can think of a quantum system that corresponds to the ALG with OBC by setting $U_1 = 2 J$ and $U_2 = \varepsilon J$ and adding a boundary term: $W_{\mathcal{C}, \mathcal{C}'} = - \braket{\mathcal{C} | H + H_\mathrm{bd} | \mathcal{C}'}$ with  $H_\mathrm{bd} := -J \hat{P} [ \sum_{i \in \partial \Omega \backslash \partial \partial \Omega} \hat{n}_i + 2 \sum_{i \in \partial \partial \Omega} \hat{n}_i + \varepsilon \sum_{j = 1}^{L_y} ( \hat{m}_{L_x \hat{x} + j \hat{y}} - \hat{m}_{1 \hat{x} + j \hat{y}} ) ] \hat{P}$.
Here we denoted the boundary points of $\Omega$ as $\partial \Omega$ and the four corner points as $\partial \partial \Omega$.

\begin{figure*}
\centering
\includegraphics[scale=0.8]{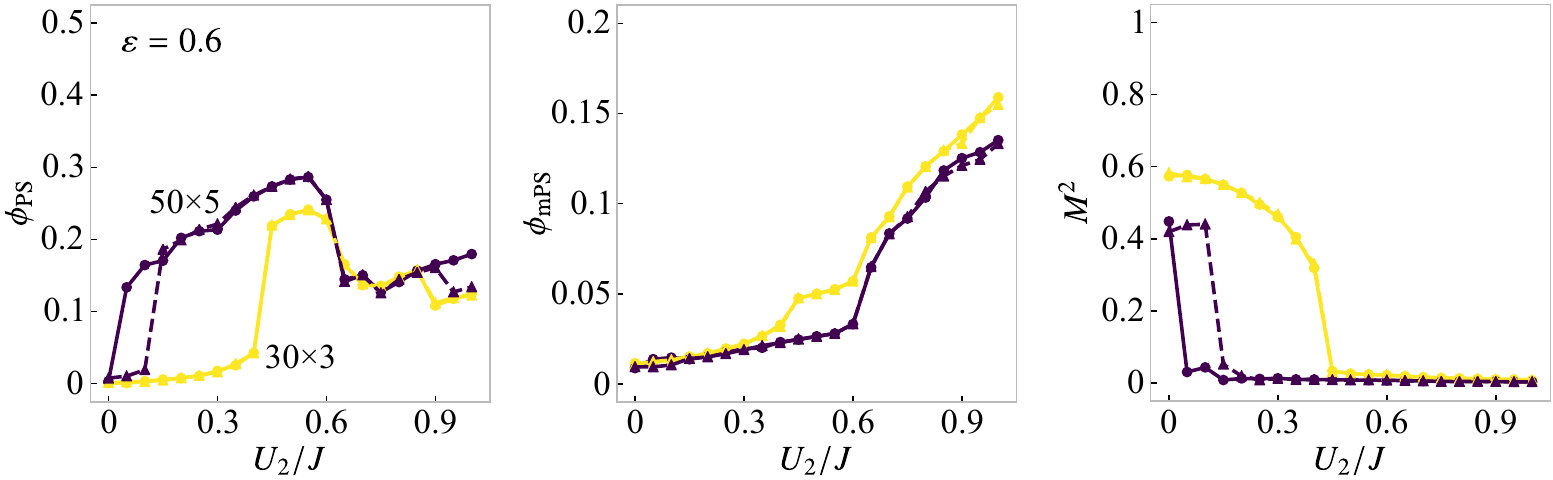}
\caption{$U_2$-dependence of $\phi_\mathrm{PS}$, $\phi_\mathrm{mPS}$, and $M^2$.
We used $\rho = 0.5$, $\varepsilon = 0.6$, $h = 0.025 J$, and $U_1 = 2 J$ in $30 \times 3$ and $50 \times 5$ systems, with the PS (solid line with circles) or the polar (dashed line with triangles) initial state.
Simulation parameters are $\Delta t = 1 / [N ( 4 J + h )]$, $N_\mathrm{c} = 10^5$, and $M = 10^4 N$.}
\label{Fig:U2dep}
\end{figure*}

\subsection{Correspondence to the ferromagnetic XXZ model}
\label{Sec:corrXXZ}

We consider the case where $\varepsilon = 0$ and $U_2 = 0$.
Since there is no spin-dependence in this model, it is equivalent to the single-component hard-core boson model ($J > 0$ and $U_1 > 0$):
\begin{equation}
H_\mathrm{HCB} = - J \sum_{\braket{i, j}} \left( a_i^\dag a_j + a_j^\dag a_i \right) - U_1 \sum_{\braket{i, j}} \hat{n}_i \hat{n}_j + \mathrm{const.}
\end{equation}
Mapping the Fock bases to spin-$1/2$ bases as $\ket{n_j = 0} \to \ket{s_j^z = - 1 / 2}$ and $\ket{n_j = 1} \to \ket{s_j^z = + 1 / 2}$, or equivalently, $a_j \to \hat{S}_j^-$ and $a_j^\dag \to \hat{S}_j^+$ with $\hat{S}_j^\pm := \hat{S}_j^x \pm i \hat{S}_j^y$, we obtain
\begin{equation}
H_\mathrm{HCB} \to H_\mathrm{XXZ} = - \sum_{\braket{i, j}} \left[ 2 J \left( \hat{S}_i^x \hat{S}_j^x + \hat{S}_i^y \hat{S}_j^y \right) + U_1 \hat{S}_i^z \hat{S}_j^z \right] + \mathrm{const.}
\end{equation}
For $U_1 > 0$, $H_\mathrm{XXZ}$ represents the ferromagnetic XXZ model.
Here, the total particle number $N$ and the system size $L_x L_y$ in the hard-core boson model are related to the total magnetization $M_\mathrm{tot}^z$ in the XXZ model as $M_\mathrm{tot}^z = N - L_x L_y / 2$.
In particular, when $U_1 = 2 J$, $H_\mathrm{XXZ}$ is nothing but the ferromagnetic Heisenberg Hamiltonian~\cite{Matsubara1956}.

\begin{figure*}
\centering
\includegraphics[scale=0.8]{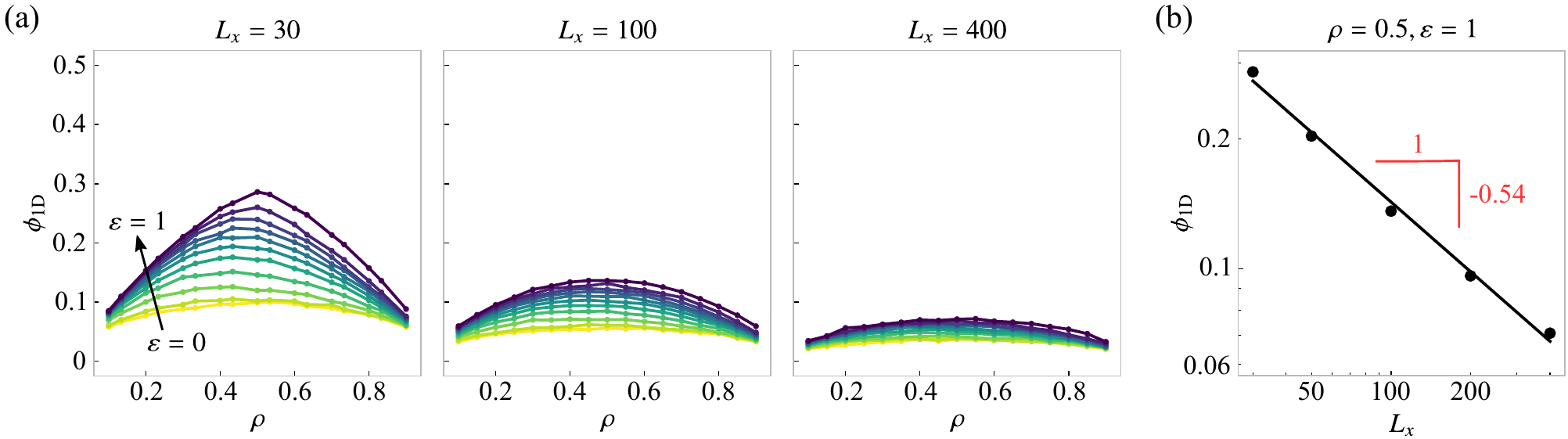}
\caption{
(a) $\rho$-dependence of $\phi_\mathrm{1D}$ for different values of $\varepsilon$ (in increments of $0.1$) in the 1D ALG with $L_x = 30, 100, 400$.
(b) $L_x$-dependence of $\phi_\mathrm{1D}$ for $\rho = 0.5$ and $\varepsilon = 1$, which indicates $\phi_\mathrm{1D} \sim {L_x}^{-0.54}$ for large $L_x$.
In both (a) and (b), we used $h = 0.05 J$.
Also, we used $\Delta t = 1 / [N ( 2 J + h )]$ and took $10^4$ samples with $M = 10^4 N$ for $L_x = 30, 50, 100$; $5 \times 10^3$ samples with $M = 2 \times 10^4 N$ for $L_x = 200$; and $3 \times 10^3$ samples with $M = 5 \times 10^4 N$ for $L_x = 400$.}
\label{Fig:1D-C}
\end{figure*}

\begin{figure*}
\centering
\includegraphics[scale=0.8]{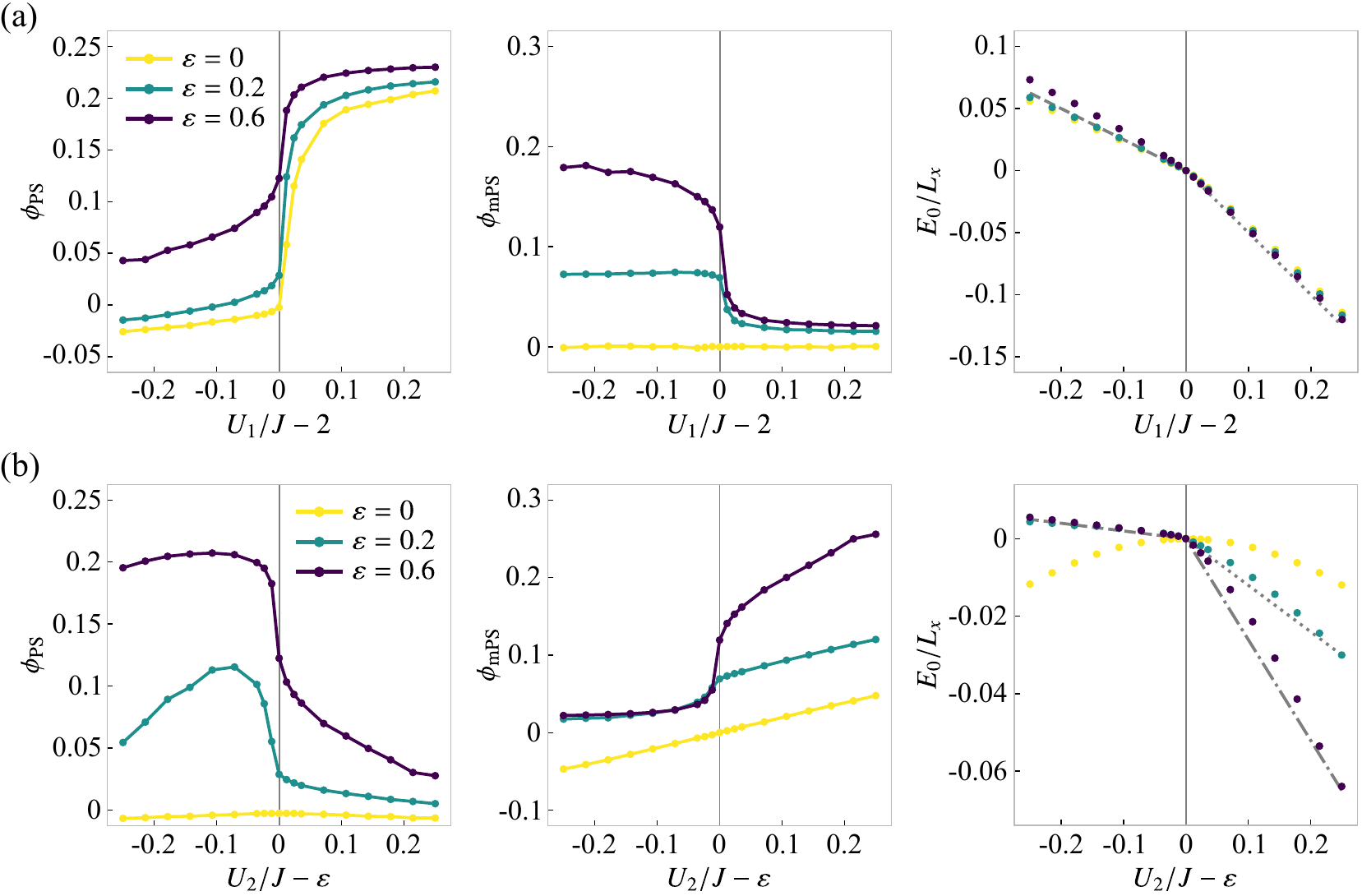}
\caption{
(a) $U_1$-dependence with $U_2 = \varepsilon J$ and (b) $U_2$-dependence with $U_1 = 2 J$ of $\phi_\mathrm{PS}$, $\phi_\mathrm{mPS}$, and $E_0$ for $\rho = 0.5$, $h = 0.025 J$, and $\varepsilon = 0, 0.2, 0.6$ in 1D systems ($L_x = 100$).
The classical condition is indicated with the gray vertical line.
For reference, in the figures of $E_0$, we also plotted $\braket{H}_\mathrm{C}$ for (a) the disordered state with no spatial correlation (dashed) and the fPS state (dotted) or (b) the fpmPS states with $N_\mathrm{cl} = 1$ (dashed), $N_\mathrm{cl} = 6$ (dotted), and $N_\mathrm{cl} = 13$ (dash-dotted).
We set $\Delta t = 1 / [N ( 2 J + h )]$ and took $N_\mathrm{c} = 10^4$ and $M = 2 \times 10^5 N$ for (a), while $N_\mathrm{c} = 10^5$ and $M = 10^4 N$ for (b).}
\label{Fig:1D-Q}
\end{figure*}

\begin{figure*}
\centering
\includegraphics[scale=0.8]{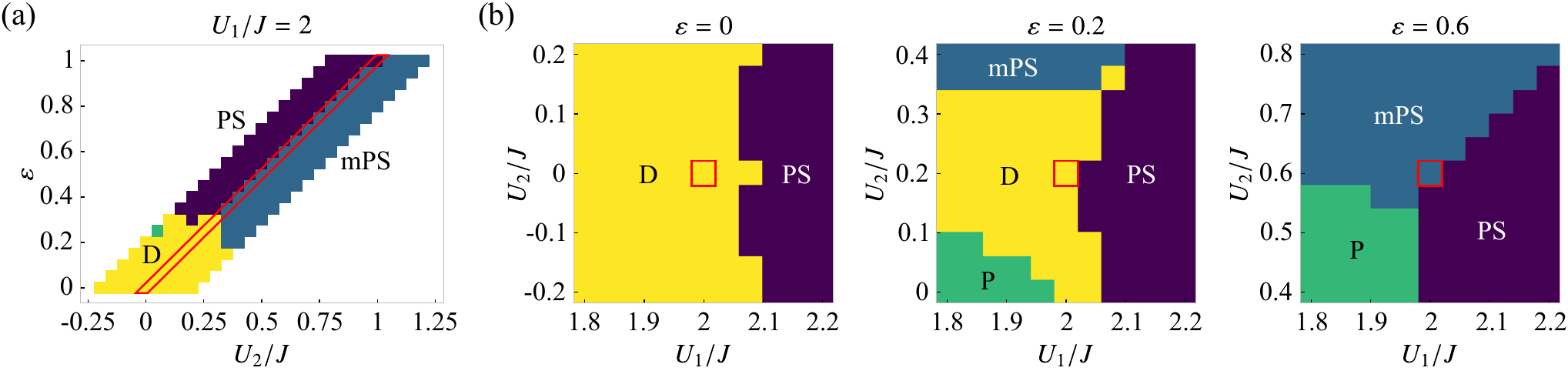}
\caption{Ground-state phase diagrams of the 1D quantum model.
(a) $U_2$-$\varepsilon$ phase diagram for $U_1 = 2 J$ around the classical line (red box) with PS ($\phi_\mathrm{PS} > 0.1$ and $\phi_\mathrm{mPS} \leq 0.1$), mPS ($\phi_\mathrm{mPS} > 0.1$), P (polar, $M^2 > 0.1$), and D (disordered, otherwise) states.
(b) $U_1$-$U_2$ phase diagrams for $\varepsilon = 0, 0.2, 0.6$ around the cross section of the classical line (red box).
In all figures, we set $\rho = 0.5$ and $h = 0.025 J$.
Simulation parameters are $\Delta t = 1 / [N ( 2 J + h )]$, $N_\mathrm{c} = 10^4$, and $M = 10^4 N$.}
\label{Fig:1D-PD}
\end{figure*}

\subsection{Energy of different states}
\label{Sec:Energy}

For an arbitrary state $\ket{\psi} = \sum_{\mathcal{C}} P (\mathcal{C}) \ket{\mathcal{C}}$, where $\ket{\mathcal{C}}$ is the Fock-space basis, we can calculate $\braket{H}_\mathrm{C}$ as
\begin{align}
& \braket{H}_\mathrm{C} = \sum_{\mathcal{C}} \left[ \left( 2 J - U_1 \right) \sum_{\braket{i, j}} n_i n_j \right. \nonumber \\
& \left. + \left( \varepsilon J - U_2 \right) \sum_i m_i \left( n_{i + \hat{x}} - n_{i - \hat{x}} \right) \right] P ( \mathcal{C} ) \Bigg/ \sum_\mathcal{C} P ( \mathcal{C} ).
\end{align}
Here, $n_i$ and $m_i$ are the local density and magnetization for the configuration $\mathcal{C}$, respectively.

In Fig.~\ref{Fig:QMIPS}, we plotted $\braket{H}_\mathrm{C}$ calculated for the disordered state with no spatial correlation, the fully phase-separated (fPS) state, and the fully polarized microphase-separated (fpmPS) state (Fig.~\ref{Fig:RefStates}).
First, the disordered state with no spatial correlation is defined as $\ket{\psi} = ( \sum_{i, s} a_{i, s}^\dag )^N \ket{0}$, and the corresponding energy is $\braket{H}_\mathrm{C} = 2 ( 2 J - U_1 ) \rho^2 L_x L_y$ by neglecting $o ( L_x L_y )$, which we plot in Fig.~\ref{Fig:QMIPS}(a) (dashed line).
Second, we define a fPS state as $\ket{\psi} = \prod_{i \in \Omega} ( a_{i, +}^\dag + a_{i, -}^\dag ) \ket{0}$, where $\Omega$ is an area containing $N$ sites and minimizing the circumference [Fig.~\ref{Fig:RefStates}(a)].
The corresponding energy is $\braket{H}_\mathrm{C} = 2 ( 2 J - U_1 ) \rho L_x L_y$ by neglecting $o ( L_x L_y )$, which we plot in Fig.~\ref{Fig:QMIPS}(a) (dotted line).
Lastly, we define a fpmPS state with $N_\mathrm{cl}$ clusters, assuming commensurability, as ${\ket{\psi}} = \prod_{n = 1}^{N_\mathrm{cl}} [ \prod_{i \in \Omega_n \setminus ( \partial \Omega_n^\mathrm{L} \cup \partial \Omega_n^\mathrm{R} )} ( a_{i, +}^\dag + a_{i, -}^\dag ) \prod_{i \in \partial \Omega_n^\mathrm{L}} a_{i, +}^\dag \prod_{i \in \partial \Omega_n^\mathrm{R}} a_{i, -}^\dag ] \ket{0}$, where $\Omega_n$ is the $n$-th rectangular area and $\partial \Omega_n^\mathrm{L (R)}$ is its left (right) boundary [Fig.~\ref{Fig:RefStates}(b)].
The corresponding energy is $\braket{H}_\mathrm{C} = ( 2 J - U_1 ) ( 2 \rho L_x L_y - N_\mathrm{cl} L_y ) + 2 ( \varepsilon J - U_2 ) N_\mathrm{cl} L_y$, which we plot with $N_\mathrm{cl} = 1$ (dashed line) and with $N_\mathrm{cl} = 4$ (dotted line) in Fig.~\ref{Fig:QMIPS}(b).
Note that, for $U_2 \gg J, h, U_1$ ($> 0$), a fpmPS state with $N_\mathrm{cl} = \rho L_x / 2$ [Fig.~\ref{Fig:RefStates}(c)] is the ground state within the first-order perturbation of $h / U_2$, $J / U_2$, and $U_1 / U_2$.

\begin{figure*}
\centering
\includegraphics[scale=0.8]{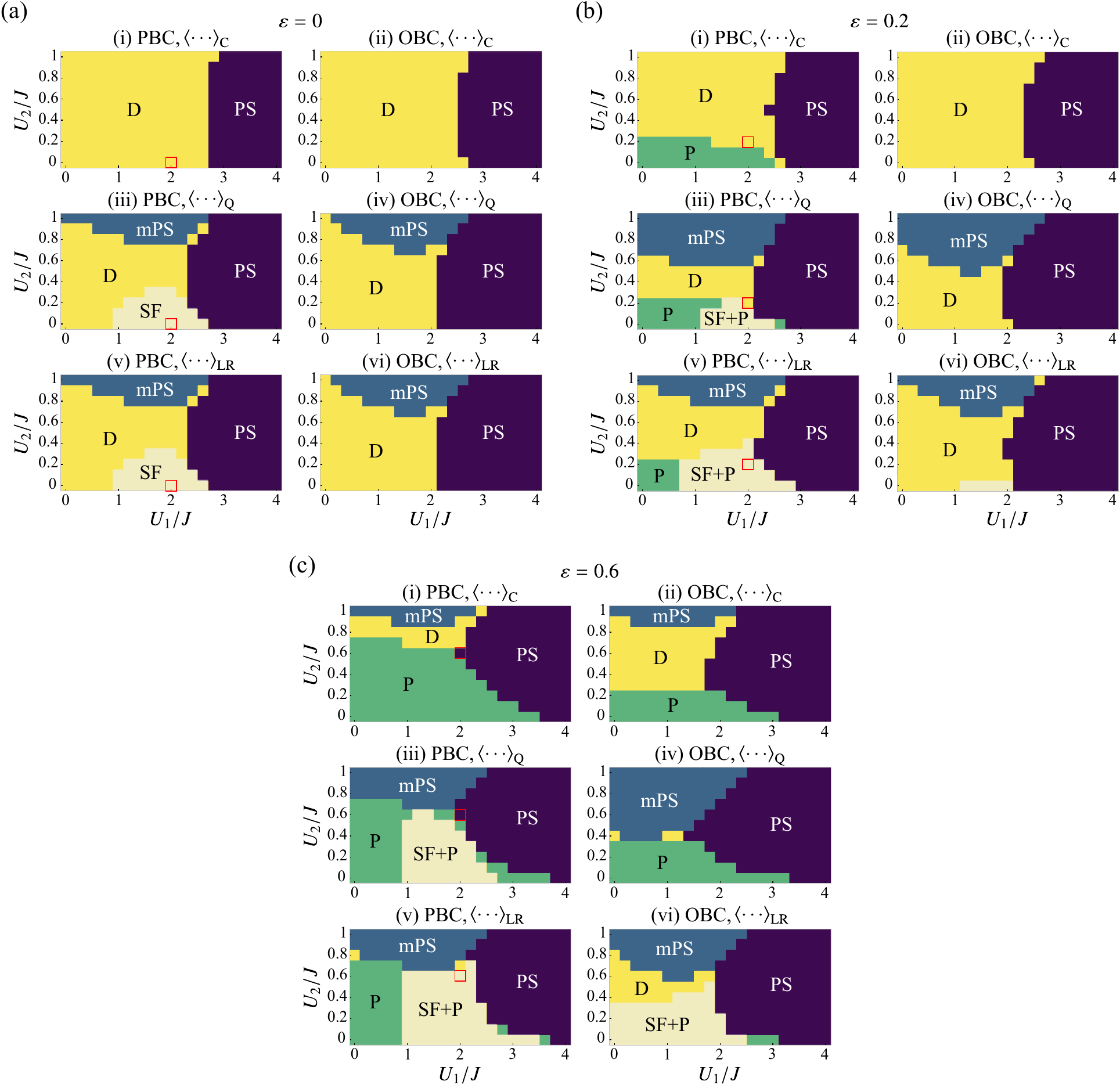}
\caption{$U_1$-$U_2$ phase diagrams in small 1D systems.
The used parameters are $L_x = 12$, and  (a) $\varepsilon = 0$, (b) $0.2$, and (c) $0.6$, respectively, with PS ($\phi_\mathrm{PS} > 0.05$ and $\phi_\mathrm{mPS} \leq 0.3$), mPS ($\phi_\mathrm{mPS} > 0.3$), P (polar, $M^2 > 0.2$), SF (superfluid, $\phi_\mathrm{SF} > 0.2$), and D (disordered, otherwise) states.
The order parameters are calculated by exact diagonalization, using (i, ii) $\braket{\cdots}_\mathrm{C}$, (iii, iv) $\braket{\cdots}_\mathrm{Q}$, or (v, vi) $\braket{\cdots}_\mathrm{LR}$, for the PBC (i, iii, v) or OBC (ii, iv, vi).}
\label{Fig:ED}
\end{figure*}

\subsection{Convergence of simulations and asymmetric-hopping-induced phase separation}
\label{Sec:convergence}

In the diffusion Monte Carlo (DMC) simulations, we checked the convergence to the steady-state by examining the initial-state dependence of the results and the relaxation of the order parameters and the ground-state energy.
As an illustration, we show the $U_1$-dependence of $\phi_\mathrm{PS}$ and $\phi_\mathrm{mPS}$ obtained with the fPS initial state, compared with that obtained with the disordered initial state [Fig.~\ref{Fig:Conv}(a) and also see ~\ref{Fig:QMIPS}(a) in the main text].
Apart from statistical errors, we do not see differences due to initial conditions for the case of system size $30 \times 3$, but there is a discrepancy in the case of $50 \times 5$ (see also Fig.~\ref{Fig:U2dep}).
This is likely due to the number of clones being insufficient for the large system size simulation~\cite{nemoto2016}.
Further, we show an example of the time-dependence of $\phi_\mathrm{PS}$ and $E_0$ evolving from the disordered initial state in the DMC simulations [Fig.~\ref{Fig:Conv}(b)], which indicates that the steady-state is achieved in the final state.
Note that, for $U_1 = 2 J$ and $U_2 = \varepsilon J$ (classical condition), $E_0$ is trivially zero according to the probability conservation.

We show the $U_2$-dependence of the order parameters for $30 \times 3$ and $50 \times 5$ systems (Fig.~\ref{Fig:U2dep}).
We can see that the polar state with finite $M^2$ is destabilized and instead the PS state with finite $\phi_\mathrm{PS}$ dominates broader parameter regions as the system becomes larger, though the dependence on the initial state remains around the phase boundary in the $50 \times 5$ system.
Thus, the PS state may replace the polar state even for $U_2 = 0$ in larger systems, and thus the asymmetric-hopping-induced phase separation can occur as observed in 1D systems [Fig.~\ref{Fig:ep}(b) in the main text].

\subsection{1D model}
\label{Sec:1Dmodel}

For the 1D counterpart of the ALG, we show the $\rho$ and $\varepsilon$-dependence [Fig.~\ref{Fig:1D-C}(a)] and the size-dependence [Fig.~\ref{Fig:1D-C}(b)] of the order parameter of phase separation, $\phi_\mathrm{1D}$ [$:= -\min C(r)$], where $C (r)$ [$:= {L_x}^{-1} \sum_i \braket{[n (x_i + r) - \rho] [n (x_i) - \rho]}$] is the 1D density correlation function.
The data suggest that the macroscopic MIPS is not stable in the thermodynamic limit ($\phi_\mathrm{1D} \to 0$ for $L_x \to \infty$).
This result is consistent with preceding studies of similar 1D models~\cite{Thompson2011, Soto2014, Cates2015}, where the macroscopic MIPS does not occur due to the spontaneous formation of domain boundaries.

For the quantum model, Figs.~\ref{Fig:1D-Q} and \ref{Fig:1D-PD} show the 1D counterparts of Figs.~\ref{Fig:QMIPS} and \ref{Fig:PD} in the main text, respectively.
We can see that the discontinuous transition occurs in crossing the classical line (Fig.~\ref{Fig:1D-Q}) as observed in 2D systems, and the topology of the phase diagrams (Fig.~\ref{Fig:1D-PD}) is also similar.
Note that in 1D systems with finite $\varepsilon$ or $U_2$, the mPS order parameter $\phi_\mathrm{mPS}$ [$:= {L_x}^{-1} \sum_{i = 1}^{L_x} \braket{\hat{m}_{i} ( \hat{n}_{i + 1} - \hat{n}_{i - 1} )}_\mathrm{C}$] is generically non-zero even for $L_x \to \infty$, and consequently the disordered and mPS states are indistinguishable from the symmetry perspective.

\section{Quantum phase diagrams in small 1D systems}
\label{Sec:ED}

To clarify how the phase diagrams depend on the definition of order parameters and the boundary condition, we calculated the order parameters using exact diagonalization in small 1D systems. On top of the expectation values defined in the main text, we consider $\braket{\cdots}_\mathrm{LR} = \braket{\psi'_0 | \cdots | \psi_0}$ with $\bra{\psi'_0}$ being the left ground state. We additionally define the order parameters, $\phi_\mathrm{PS}$, $\phi_\mathrm{mPS}$, $M^2$, and $\phi_\mathrm{SF}$ for $\braket{\cdots}_\mathrm{LR}$. 

The results are summarized in the phase diagrams (Fig.~\ref{Fig:ED}). First, we find that all the states predicted using $\braket{\cdots}_\mathrm{C}$ with the PBC [(i) in Figs.~\ref{Fig:ED}(a)-(c)] appear, regardless of the definition of order parameters or the boundary condition.
Thus, the DMC simulation, which is applicable to larger systems as demonstrated in the main text, is useful in qualitatively predicting the phase diagram (apart from the SF order) in the experimentally relevant case, where we use $\braket{\cdots}_\mathrm{Q}$ with the OBC.
Next, focusing on the cases with the PBC, we see that the SF state appears for $\varepsilon = 0$ [(iii, v) in Fig.~\ref{Fig:ED}(a)] consistently with the previous studies of the Hermitian hard-core boson models~\cite{Matsubara1956}.
Interestingly, the SF state with polar order is stable for finite $\varepsilon$ [(iii, v) in Figs.~\ref{Fig:ED}(b) and (c)].
Lastly, since the OBC prevents the particles from flowing, the polar order is suppressed [(ii, iv, vi) in Fig.~\ref{Fig:ED}(b)] unless $\varepsilon$ is large enough [(ii, iv, vi) in Fig.~\ref{Fig:ED}(c)].

\bibliographystyle{apsrev4-2}
\bibliography{reference}

\end{document}